\documentstyle[12pt,eqsection]{article}
\newcommand{\be}{\begin{equation}}
\newcommand{\ee}{\end{equation}}
\newcommand{\bea}{\begin{eqnarray}}
\newcommand{\eea}{\end{eqnarray}}
\newcommand{\norsl}{\normalsize\sl}
\newcommand{\norsc}{\normalsize\sc}

\def\Av{\mbox{\boldmath $A$}}

\def\Cv{\mbox{\boldmath $C$}}

\def\Kv{\mbox{\boldmath $K$}}

\def\Xv{\mbox{\boldmath $X$}}

\def\qv{\mbox{\boldmath $q$}}
\def\kv{\mbox{\boldmath $k$}}

\begin{document}

\begin{titlepage}

\title{ Spin Structure Function of the  Virtual Photon
\\ Beyond the Leading Order in QCD}
\author{
\norsc  Ken SASAKI\thanks{e-mail address: sasaki@ed.ynu.ac.jp}~ and
      Tsuneo UEMATSU\thanks{e-mail address: uematsu@phys.h.kyoto-u.ac.jp} \\
\norsl  Dept. of Physics,  Faculty of Engineering, Yokohama National
University \\
\norsl  Yokohama 240-8501, JAPAN \\
\norsl  Dept. of Fundamental Sciences, FIHS, Kyoto University \\
\norsl     Kyoto 606-8501, JAPAN \\
}

\date{}
\maketitle

\begin{abstract}
{\normalsize
Polarized photon structure can be studied in the future polarized $e^{+}e^{-}$
colliding-beam experiments.
We investigate the spin-dependent structure function of the virtual photon
$g_1^{\gamma}(x,Q^2,P^2)$, in perturbative QCD for $\Lambda^2 \ll P^2 \ll
Q^2$, where $-Q^2$  ($-P^2$) is the mass squared of the probe (target)
photon.
The analysis is performed to next-to-leading order in QCD.
We particularly emphasize the renormalization scheme independence of the
result.The non-leading corrections significantly modify the leading log
result,
in particular, at large $x$ as well as at small $x$. We also discuss the
non-vanishing first moment sum rule of $g_1^\gamma$, where ${\cal
O}(\alpha_s)$ corrections are computed.
}
\end{abstract}

\begin{picture}(5,2)(-290,-500)
\put(2.3,-75){YNU-HEPTh-98-103}
\put(2.3,-90){KUCP-125}
\put(2.3,-105){December 1998}
\end{picture}

\thispagestyle{empty}
\end{titlepage}
\setcounter{page}{1}
\baselineskip 18pt
\section{Introduction}
\smallskip

In recent years there has been growing interest in the study of a polarized
photon structure function. The information on the spin structure of the photon
would be provided by the resolved photon process in polarized electron and
proton collision in the polarized version of HERA \cite{Barber,SVZ}. 
More directly, the spin-dependent structure function of photon $g_1^\gamma$ 
can be measured by the polarized e$^+$ e$^{-}$ collision in the future 
linear colliders (Fig.1).

{}From the theoretical viewpoint, the first moment of a photon structure
function $g_1^\gamma$ has recently attracted attention 
in the literatures \cite{BASS,ET,NSV,FS,BBS} in
connection with its relevance for the axial anomaly, which has also played
an important role in the QCD analysis of the spin structure of the
nucleon. Our aim here
is to carry out the QCD computation of photon's polarized structure function
at the same level of unpolarized case.
Here we note that the two-loop splitting functions of DGLAP equation or
equivalently the two-loop anomalous dimensions have recently been
calculated \cite{MvN,V}, and we can perform the next-to-leading order
QCD analysis for the polarized photon structure function. Actually there
has already been an analysis of spin-dependent structure function $g_1^\gamma$
for the real photon target by Stratmann and Vogelsang \cite{SV}.

In this paper we shall investigate the polarized virtual photon structure
function $g_1^\gamma(x,Q^2,P^2)$ to the next-leading order (NLO) in QCD,
in the kinematical region:
\be
\Lambda^2 \ll P^2 \ll Q^2
\ee
where $-Q^2$ ($-P^2$) is the mass squared of the probe (target) photon,
and $\Lambda$ is the QCD scale parameter.
We can base our arguments either on DGLAP-type $Q^2$ evolution equation for
the parton distributions or on the framework of operator product expansion
(OPE) and the renormalization group (RG) method.
The unpolarized virtual photon structure functions $F_2^\gamma(x,Q^2,P^2)$
and $F_L^\gamma(x,Q^2,P^2)$ were studied in the leading order (LO) \cite{UW1}
and in the NLO \cite{UW2,ROS}. And the parton contents of virtual photon
were studied in ref.\cite{DG,GRS} and the target mass effect of unpolarized and
polarized virtual photon structure in LO was discussed in ref.\cite{MR}.

The advantage to study the virtual photon target is that we can calculate
the whole structure function entirely up to NLO by the perturbative method.
On the other hand, for the real photon target \cite{BB,DO}
we can calculate the 
perturbative pieces, but not the non-perturbative contributions which
may be estimated, for example, by vector-dominance model \cite{EW}.
The perturbative pieces for the real photon target can be reproduced
from the result for the virtual photon case.

In the next section we discuss the polarized photon structure functions.
Next we present the two theoretical frameworks based on OPE (section 3)
and on DGLAP parton model approach (section 4). In section 5, the
sum rule for the first moment of $g_1^\gamma$ will be evaluated
up to the order of $\alpha_s$. The numerical analysis of $g_1^\gamma$ 
will be given in section 6. The final section is devoted to the conclusion
and discussion. 

\section{Polarized photon structure functions}
\smallskip

Let us consider the forward virtual photon scattering amplitude (Fig.2),
\be
T_{\mu\nu\rho\tau}(p,q)=i\int d^4 x d^4 y d^4 z e^{iq\cdot x}e^{ip\cdot (y-z)}
\langle 0|T(J_\mu(x) J_\nu(0) J_\rho(y) J_\tau(z))|0\rangle.
\ee
Its absorptive part is related to the structure tensor 
$W_{\mu\nu\rho\tau}(p,q)$ for the photon with mass squared 
$p^2=-P^2$ probed by the photon with $q^2=-Q^2$:
\be
W_{\mu\nu\rho\tau}(p,q)=\frac{1}{\pi}{\rm Im}T_{\mu\nu\rho\tau}(p,q)~.
\ee
The anti-symmetric part, $W_{\mu\nu\rho\tau}^{A}$, which is
antisymmetric under the interchange of $\mu$ and $\nu$, can be decomposed as
\bea
W_{\mu\nu\rho\tau}^{A}&=&
\epsilon_{\mu\nu\lambda\sigma}q^\lambda
{\epsilon_{\rho\tau}}^{\sigma\beta}p_\beta \frac{1}{p\cdot
q}g_1^\gamma\nonumber\\
&+&
\epsilon_{\mu\nu\lambda\sigma}q^\lambda
(p\cdot q\ {\epsilon_{\rho\tau}}^{\sigma\beta}p_\beta-\epsilon_{\rho\tau
\alpha\beta}p^\beta p^\sigma q^\alpha )\frac{1}{(p\cdot q)^2}g_2^\gamma~,
\eea
which gives two spin-dependent structure functions, $g_1^\gamma(x,Q^2,P^2)$ 
and $g_2^\gamma(x,Q^2,P^2)$. For a real photon, $g_2^\gamma$ is identically 
zero, and there exists only one spin struture function, $g_1^\gamma(x,Q^2)$. 
On the other hand, for the off-shell or virtual photon ($P^2 \neq 0$) target, 
we have two 
spin-dependent structure functions $g_1^\gamma$ and $g_2^\gamma$.
More detailed argument on the structure functions is given in the Appendix D.
The $g_1^\gamma$ is related to the structure function $W_4^\gamma$, which was
discussed some years ago in \cite{BM,AMR}, such that $g_1^\gamma(x,Q^2)\equiv
2W_4^\gamma(x,Q^2)$. Here we note that the LO QCD correction
was first studied by one of the authors in \cite{KS} and later in 
\cite{MANO,BASS}.

First we note that the same framework used in the analysis of the nucleon spin
structure functions can be applied in our case.  We can either base our 
argument on the OPE supplementd by the RG method, or on the DGLAP type parton 
evolution equations. It should be noted the next-to-leading order analysis is 
now possible since the two-loop anomalous dimensions of twist-2 operators 
$R_n^i$ relevant for the spin-dependent structure function (or equivalently 
two-loop parton splitting functions) were calculated
independently by two groups, by Mertig-van Neerven \cite{MvN} and by
Vogelsang \cite{V}.

\section{Theoretical framework based on OPE}
\smallskip

In our previous paper \cite{SU}, we based our argument on the QCD improved
parton model approach. Here we start with theoretical framework based on the 
OPE and RG method. Applying OPE for the product of two electromagnetic 
currents  we get
\bea
&&\hspace{-1cm}i\int d^4x e^{iq\cdot x}T(J_\mu(x)J_\nu(0))^A=
-i\varepsilon_{\mu\nu\lambda\sigma}q^\lambda
\sum_{n=odd}\left(\frac{2}{Q^2}\right)^n \hspace{-0.3cm}
q_{\mu_1}\cdots q_{\mu_{n-1}}
\sum_i C^i_{1,n} R_{1,i}^{\sigma\mu_1\cdots q_{\mu_{n-1}}} \nonumber\\
&&\hspace{-1cm}-i(\varepsilon_{\mu\rho\lambda\sigma}q_\nu q^\rho - 
\varepsilon_{\nu\rho\lambda\sigma}q_\mu q^\rho
-q^2\varepsilon_{\mu\nu\lambda\sigma})
\sum_{n=odd}
\left(\frac{2}{Q^2}\right)^n \hspace{-0.3cm}
q_{\mu_1}\cdots q_{\mu_{n-2}}
\sum_i C^i_{2,n} R_{2,i}^{\lambda\sigma\mu_1\cdots q_{\mu_{n-2}}}.
\eea
For polarized deep inelastic scattering, the twist-2 and twist-3 operators:
$R_1^n, \ R_2^n$
contribute to the structure functions in the scaling limit. For $g_1^\gamma$
only twist-2 operators are relevant.
Now we can write down the moment sum rule for $g_1^\gamma$:
\be
\int_0^1dx x^{n-1} g_1^\gamma(x,Q^2,P^2)=
\sum_{i=\psi,G,NS,\gamma} C_n^i(Q^2/\mu^2,{\bar g}(\mu^2),\alpha)
\langle\gamma(p)|R_n^i(\mu^2)|\gamma(p)\rangle ~,
\label{moments}
\ee
where $R_n^i$ and $C_n^i$ are the twist-2 operators and their
coefficient functions (Hereafter we suppress the index 1 for twist-2
operators),
with $\mu$ being the renormalization point and
$\alpha=e^2/4\pi$, the QED coupling constant. $\psi$, $G$, $NS$
and $\gamma$ stand for singlet quark, gluon, non-singlet quark and
photon, respectively. The relevant twist-2 operators $R_n^i$ 
($i=\psi(S),G,NS,\gamma$) are given by \cite{KS}:
\bea
  R_{\psi}^{\sigma\mu_{1}\cdots \mu_{n-1}} &=&
         i^{n-1}  \overline{\psi}
       \gamma^{\{\sigma}D^{\mu_1} \cdots D^{\mu_{n-1}\}}\gamma_5~1\psi -
            {\rm trace \ terms}
, \label{quark}\\
  R_G^{\sigma\mu_{1}\cdots \mu_{n-1}} &=& \frac{1}{4}~i^{n-1}
        \epsilon^{\{\sigma}_{\ \ \alpha\beta\gamma}G^{\alpha\mu_1}
               D^{\mu_2} \cdots D^{\mu_{n-1}\}}G^{\beta\gamma} -
            {\rm trace \ terms}
     \label{gluon}\\
  R_{NS}^{\sigma\mu_{1}\cdots \mu_{n-1}} &=&
         i^{n-1}  \overline{\psi}
       \gamma^{\{\sigma}D^{\mu_1} \cdots D^{\mu_{n-1}\}}\gamma_5
             (Q^2_{ch}-1)\psi  -
            {\rm trace \ terms} \\
 R_{\gamma}^{\sigma\mu_{1}\cdots \mu_{n-1}} &=& \frac{1}{4}~i^{n-1}
        \epsilon^{\{\sigma}_{\ \ \alpha\beta\gamma}F^{\alpha\mu_1}
               \partial^{\mu_2} \cdots \partial^{\mu_{n-1}\}}F^{\beta\gamma}
                - {\rm trace \ terms}
     \label{photon}
\eea
where $\{\ \ \  \}$ means complete symmetrization over the Lorentz indices
$\sigma\mu_{1}\cdots \mu_{n-1}$, $D^{\mu}$ denotes covariant derivative,
$1$ is an $n_f \times n_f$ unit matrix and $Q^2_{ch}$ is the square of the
$n_f \times n_f$ quark-charge matrix, with $n_f$ being the number of flavors.
Here we note that the essential feature in the above equation
is the appearance of photon operators $R_n^\gamma$ in addition to the 
hadronic operators.

For $-p^2=P^2 \gg \Lambda^2$, we can calculate the photon matrix
elements of the hadronic operators perturbatively.
Choosing $\mu^2$ to be close to $P^2$, we get, to the lowest order,
\be
\langle\gamma(p)|R_n^i(\mu)|\gamma(p)\rangle
=\frac{\alpha}{4\pi}\left(-\frac{1}{2}K_n^{0,i}\ln\frac{P^2}{\mu^2}
+A_n^{i}\right), \quad i=\psi(S),G,NS
\ee
where $K_n^{0,i}=(\Kv_n^0)^i$ are one-loop anomalous dimension
matrix elements between the photon and hadronic operators.
On the other hand, in the leading order of the QED coupling constant,
$\alpha$, we have for the photon operator $R_n^\gamma$:
\be
\langle\gamma(p)|R_n^\gamma(\mu)|\gamma(p)\rangle=1
\ee 

It should be noted that the finite term $A_n^{i}$ depends on the 
renormalization scheme for the operators $R_n^i$.
Putting $\mu^2=-p^2=P^2$, we have
\be
\langle\gamma(p)|R_n^i(\mu)|\gamma(p)\rangle|_{\mu^2=P^2}=
\frac{\alpha}{4\pi}A_n^{i}~,
\ee
and the $n$-th moment with this choice $\mu^2=P^2$ in (\ref{moments}) becomes
\bea
&&\int_0^1 dx x^{n-1}g_1^\gamma(x,Q^2,P^2)\\
&&=
\sum_{i,j=\psi,G,NS,\gamma}\langle\gamma(p)|R_n^i(\mu^2=P^2)|\gamma(p)\rangle
\left(T\exp\left[\int_{{\bar g}(Q^2)}^{{\bar g}(P^2)}dg
\frac{\gamma_n(g)}{\beta(g)}
\right]\right)_{ij} C_n^j(1,{\bar g},\alpha)~.\nonumber
\eea
The evolution factor in the last equation is found to be \cite{BB}:
\be
T\exp\left[\int_{{\bar g}(Q^2)}^{{\bar g}(P^2)}dg\frac{\gamma_n(g)}{\beta(g)}
\right]
=
\left(
\begin{array}{c|c}
M_n & {\bf 0} \\
\hline
\Xv_n & {\bf 1}
\end{array}
\right)
\ee
where
\bea
M_n(Q^2/P^2,{\bar g}(P^2))&=&
T\exp\left[\int_{{\bar g}(Q^2)}^{{\bar g}(P^2)}dg\frac{{\widehat\gamma}_n(g)}
{\beta(g)}\right]\nonumber\\
\Xv_n(Q^2/P^2,{\bar g}(P^2),\alpha)&=&
\int_{{\bar g}(Q^2)}^{{\bar g}(P^2)}dg\frac{\Kv_n(g,\alpha)}{\beta(g)}
T\exp\left[\int_{{\bar g}(Q^2)}^{g}dg'\frac{{\widehat\gamma}_n(g')}{\beta(g')}
\right]
\eea
with ${\widehat\gamma}_n$ and $\Kv_n$ the hadronic anomalous dimension matrix 
and
the off-diagonal element representing the mixing between the photon and hadron
operators (see Appendix A).
Thus we get
\bea
\int_0^1 dx x^{n-1}g_1^\gamma(x,Q^2,P^2)
&=&\frac{\alpha}{4\pi}\Av_n\cdot M_n(Q^2/P^2,{\bar g}(P^2))
\Cv_n(1,{\bar g}(Q^2)) \nonumber\\
&+&\Xv_n(Q^2/P^2,{\bar g}(P^2),\alpha)\cdot\Cv_n(1,{\bar g}(Q^2))
+C_n^\gamma
\eea
with 
\be
\Av_n=(A_n^{\psi},A_n^{G},A_n^{NS})~. 
\ee
The coefficient functions are given by (see Appendix C),
\bea
\Cv_n(1,{\bar g})&=&
\left(
\begin{array}{c}
C_n^\psi(1,{\bar g}) \\
C_n^G(1,{\bar g}) \\
C_n^{NS}(1,{\bar g})
\end{array}
\right)
=\left(
\begin{array}{c}
\delta_\psi(1+\frac{{\bar g}^2}{16\pi^2}B_\psi^n)\\
\delta_\psi\frac{{\bar g}^2}{16\pi^2}B_G^n \\
\delta_{NS}(1+\frac{{\bar g}^2}{16\pi^2}B_{NS}^n)
\end{array}
\right) \nonumber\\
C_n^\gamma(1,{\bar g},\alpha)&=&\frac{\alpha}{4\pi}\delta_\gamma B_\gamma^n
\eea
with $\delta_\psi=<e^2>=\sum_{i=1}^{n_f}e_i^2/n_f,  
\delta_{NS}=1, \delta_\gamma=3n_f<e^4>=3\sum_{i=1}^{n_f}e_i^4$.

We then derive the following formula for the moments:
\bea
&&\int_0^1 dx x^{n-1}g_1^\gamma(x,Q^2,P^2)  \nonumber\\
&=&\frac{\alpha}{4\pi}\frac{1}{2\beta_0}
\left[\sum_{i=+,-,NS}\tilde{P}_i^n\frac{1}{1+\lambda_i^n/2\beta_0}
\frac{4\pi}{\alpha_s(Q^2)}
\left\{1-\left(\frac{\alpha_s(Q^2)}{\alpha_s(P^2)}\right)^{\lambda_i^n/2\beta_0
+1}\right\}\right.\nonumber\\
&&\hspace{1.5cm}+\left.\sum_{i=+,-,NS}{\cal A}_i^n\left\{1-\left(\frac{\alpha_s(Q^2)}{\alpha_s(P^2)}\right)^{\lambda_i^n/2\beta_0}\right\}\right.\nonumber\\
&&\hspace{1.5cm}+\left.\sum_{i=+,-,NS}{\cal B}_i^n\left\{1-\left(\frac{\alpha_s(Q^2)}{\alpha_s(P^2)}\right)^{\lambda_i^n/2\beta_0+1}\right\}
+{\cal C}^n +{\cal O}(\alpha_s) \ \right]~.
\label{master1}
\eea
Here $\alpha_s(Q^2)={\bar g}^2(Q^2)/4\pi$ is the QCD running coupling
constant. In (\ref{master1}), we have defined
\be
{\tilde P}^n_i={\Kv}_n^0 P^n_i {\Cv}_n(1,0) \quad (i=+,-,NS).
\label{tildeP}
\ee
where $P_i^n$'s are projection operators given in the Appendix A.
The coefficients ${\cal A}_i^n$, ${\cal B}_i^n$ and ${\cal C}^n$
are computed from the NLO perturbative calculation,
and are given by
\bea
 {\cal A}^n_i &=& -{\Kv}_n^0 \sum_j \frac{P^n_j \widehat \gamma_n^{(1)} P^n_i}
             {2\beta_0 + \lambda^n_j - \lambda^n_i } {\Cv}_n(1,0)
     \frac{1}{\lambda^n_i/2\beta_0}
            -{\Kv}_n^0 \frac{\beta_1}{\beta_0}  P^n_i {\Cv}_n(1,0)
      \frac{1-\lambda^n_i/2\beta_0}{\lambda^n_i/2\beta_0} \nonumber  \\
    & & + {\Kv}_n^1 P^n_i {\Cv}_n(1,0) \frac{1}{\lambda^n_i/2\beta_0}
      - 2\beta_0 {\Av}_n P^n_i {\Cv}_n(1,0)   \label{calA}\\
\nonumber \\
  {\cal B}^n_i &=& {\Kv}_n^0 \sum_j \frac{P^n_i \widehat \gamma_n^{(1)} P^n_j}
             {2\beta_0 + \lambda^n_i - \lambda^n_j } {\Cv}_n(1,0)
     \frac{1}{1+\lambda^n_i/2\beta_0} \nonumber  \\
   &+& {\Kv}_n^0 P^n_i \pmatrix{\delta_{\psi} B^n_{\psi}  \cr
      \delta_{\psi} B^n_G \cr \delta_{NS} B^n_{NS} \cr}
    \frac{1}{1+\lambda^n_i/2\beta_0} 
     -{\Kv}_n^0 \frac{\beta_1}{\beta_0}  P^n_i {\Cv}_n(1,0)
      \frac{\lambda^n_i/2\beta_0}{1+\lambda^n_i/2\beta_0}  \\
\nonumber \\
   {\cal C}^n &=& 2\beta_0 (\delta_{\gamma}B^n_{\gamma} +
         {\Av}_n \cdot{\Cv}_n(1,0)  ) 
\label{calC}
\eea
where $\lambda_i^n$ ($i=+,-,NS$) are eigenvalues of the 1-loop anomalous 
dimension matrix ${\widehat\gamma}_n^{(0)}$ and are given in the 
Appendix A.
$\beta_0$ and the $\beta_1$ are the one- and two-loop $\beta$ functions,
and $\beta_0=11-2n_f/3$ and $\beta_1=102-38n_f/3$.

All the quantities necessary to evaluate ${\widetilde P}^n_i$,
${\cal A}_i^n$, ${\cal B}_i^n$, and ${\cal C}^n$  are now known and will be
presented in Appendix. Two-loop results \cite{MvN,V} have been calculated 
in the $\overline{\rm MS}$ scheme \cite{BBDM}.
Actually the expressions of Eqs.(\ref{master1})
and (\ref{tildeP})--(\ref{calC}) are the same in form as the ones obtained 
before by one of the authors and Walsh for the case of the virtual photon 
structure function $F_2^{\gamma}$~\cite{UW2}.
The explicit expressions of the one-loop and two-loop
anomalous dimensions \cite{MvN,V}
as well as one-loop coefficient functions \cite{KMMSU,KMSU,JK,BQ,ZvN,MvN,V} 
are given in the Appendix B and C.

Eq.(\ref{master1}) is our main result of the present paper.
The first term is the LO result, and the remaining terms
are the NLO QCD corrections. 

Now let us examine the renormalization scheme independence of the coefficients;
${\cal A}_i^n$, ${\cal B}_i^n$ and ${\cal C}^n$. As in the unpolarized case, 
${\cal B}_i^n$ can be written in terms of a scheme-independent combination of 
2-loop anomalous dimensions and 1-loop coefficient functions in the hadronic 
sector. Using the scheme-independent coefficients 
$R_{2,n}^i$ (\cite{BBL,B,GR0}), we can write
\be
{\cal B}_i^n=L_i^n R_{2,n}^i \quad (i=+,-,NS)
\ee
where the explicit form for $R_{2,n}^i$ is given in Eqs.(9)-(12) of
ref.\cite{BBL} (see also \cite{UW2}) and
\be
L_i^n={\tilde P}_i^n\frac{1}{1+\lambda_i^n/2\beta_0}
\ee
which is the coefficient of the leading-log term. The scheme independence of
${\cal B}_i^n$ follows from these two equations.

Regarding ${\cal C}^n$, we first consider the photon matrix elements of the 
renormalized quark and gluon operators.
The finite matrix elements $\Av_n$ and the tree-level coefficient
functions $\Cv_n(1,0)$ are given by
\be
\Av_n=6(<e^2>,0,<e^4>-<e^2>^2)\tilde{A}^\psi_{nG}
\ee
\be
\Cv_n(1,0)=\pmatrix{<e^2> \cr 0 \cr 1\cr}
\ee
Hence we have
\be
\Av_n\cdot\Cv_n(1,0)=6<e^4>\tilde{A}^\psi_{nG}
\ee
Noting that
\be
B^n_\gamma=\frac{2}{n_f}B^n_G, \quad \delta_\gamma=3n_f<e^4>
\ee
we find ${\cal C}^n$ equal to be
\be
{\cal C}^n=12\beta_0<e^4>(B^n_G+\tilde{A}^\psi_{nG})~.
\label{CalC}
\ee
Since the combination $B_G^n+\tilde{A}^\psi_{nG}$ is scheme
independent \cite{BBDM}, so is ${\cal C}^n$.
In fact, in the $\overline{\rm MS}$ scheme \cite{MvN}, 
the gluon matrix elements of quark operators ($i=\psi, NS$) read
\bea
\langle p,s|R_i^{\sigma\mu_1\cdots\mu_{n-1}}(\mu^2)|p,s\rangle
&=&A_i(p^2,\mu^2,g)\left[\{s^\sigma p^\mu_1 \cdots
p^{\mu_{n-1}}\}-\mbox{traces}\right]
\quad(i=\psi, NS) \nonumber\\
A_\psi&=&\frac{g^2}{16\pi^2}\left(\frac{1}{2}\gamma_{\psi G}^{(0),n}
\ln\frac{-p^2}{\mu^2}+ \widetilde A_{nG}^{\psi}\right)
\eea
where the finite matrix element $\widetilde A_{nG}^{\psi}$
is given in the parton language as \cite{MvN}:
\be
\widetilde A_{nG}^{\psi}=\int^1_0 dx~ x^{n-1} a_{S,qg}^{(1)}(x)
\ee
where
\be
a_{S,qg}^{(1)}(x)= T_f \Bigl[ (4-8x) \Bigl\{{\rm ln}\  x+ {\rm ln}(1-x)
\Bigr\}\Bigr]
\ee
with $T_f=n_f/2$. Thus we get
\be
\widetilde A_{nG}^{\psi}=2n_f \biggl[  \frac{n-1}{n(n+1)} S_1(n) +
\frac{4}{(n+1)^2}-\frac{1}{n^2}-\frac{1}{n}  \biggr]~.
\ee
Therefore, from (\ref{CalC}) we finally arrive at 
\be
{\cal C}^n=24\beta_0f<e^4>\left[
\frac{2}{n}-\frac{4}{n+1}-\frac{2}{n^2}+\frac{4}{(n+1)^2}\right]
\ee
which is consistent with the Box diagram calculation.

On the other hand, in the RG scheme adopted by Kodaira \cite{JK} which 
is the momentum subtraction scheme, we have for $n\geq 3$,
\be
\tilde{A}^\psi_{nG}=0 
\ee
with the coefficient function given by
\be
B^n_G=2n_f\ \left[
\frac{2}{n}-\frac{4}{n+1}-\frac{2}{n^2}+\frac{4}{(n+1)^2}\right]~.
\ee
For $n=1$, due to the Adler-Bell-Jackiw anomaly
\be
\tilde{A}^\psi_{n=1\ G}=-2n_f
\ee
and
\be
B^{n=1}_G=0~,
\ee
by definition, since we have no gauge-invariant gluon operator
for $n=1$.
Combining this with the result for $\tilde{A}^\psi_{nG}$, we
arrive at the same result for ${\cal C}^n$.

The scheme-independence of the remaining coefficients ${\cal A}_i^n$
follows from the above arguments on ${\cal B}_i^n$ and ${\cal C}^n$
and the physically measurable moments given in Eq.(\ref{master1}).

\section{QCD improved parton model approach}
\smallskip

We now turn to the analysis based on
the QCD improved  parton model \cite{A}
using the DGLAP parton evolution equations.

Let $q^i_{\pm}(x,Q^2,P^2)$, $G^{\gamma}_{\pm}(x,Q^2,P^2)$,
$\Gamma^{\gamma}_{\pm}(x,Q^2,P^2)$ be  quark with
$i$-flavor, gluon, and  photon distribution functions with $\pm$ helicities
of the longitudinally polarized virtual photon with mass $-P^2$ \cite{SU}.
Then the spin-dependent parton distributions are defined as
\bea
    \Delta q^i &\equiv& q^i_+ + \bar{q}^i_+ -q^i_- -\bar{q}^i_- \\
  \Delta G^{\gamma} &\equiv& G^{\gamma}_+ -G^{\gamma}_- ~,
      \qquad \Delta \Gamma^{\gamma} \equiv \Gamma^{\gamma}_+
-\Gamma^{\gamma}_- ~.
\eea
In the leading order of the electromagnetic coupling constant,
$\alpha=e^2/4\pi$,
$\Delta \Gamma^{\gamma}$ does not evolute with $Q^2$ and is set to be
$\Delta \Gamma^{\gamma}(x,Q^2,P^2)=\delta(1-x)$. The quark and gluon
distributions $\Delta q^i$ and $\Delta G^{\gamma}$ satisfy the following
evolution equations:
\bea
    \frac{d \Delta q^i(x,Q^2,P^2)}{d {\rm ln} Q^2}
 &=& \int^1_x \frac{dy}{y} \Bigl\{
    \sum_j~\Delta {\widetilde P}_{q^i q^j} (\frac{x}{y}, Q^2)~ \Delta q^j
(y,Q^2,P^2) \label{DelQi} \nonumber  \\
    & + &
          \Delta {\widetilde P}_{q G} (\frac{x}{y}, Q^2)~ \Delta G^{\gamma}
(y,Q^2,P^2)
\Bigr\}  + \Delta {\widetilde P}_{q^i \gamma} (x, Q^2,P^2),   \\
& &  \nonumber  \\
   \frac{d \Delta G^{\gamma}(x,Q^2,P^2)}{d {\rm ln} Q^2}
 &=& \int^1_x \frac{dy}{y} \Bigl\{ \Delta
   {\widetilde P}_{G q} (\frac{x}{y}, Q^2)\sum_i~ \Delta q^i (y,Q^2,P^2)
  \nonumber  \\
   &+ &
   \Delta {\widetilde P}_{G G} (\frac{x}{y}, Q^2)~ \Delta
G^{\gamma}(y,Q^2,P^2)
\Bigr\}
        + \Delta {\widetilde P}_{G \gamma} (x, Q^2,P^2),
\eea
where $\Delta {\widetilde P}_{AB}$ is a polarized splitting function of
$B$-parton to $A$-parton, defined as
$\Delta {\widetilde P}_{AB}\equiv P_{A_+ B_+} - P_{A_- B_+}$
     ($ =  P_{A_- B_-} - P_{A_+ B_-}$, due to parity conservation in QCD
and QED).

For later convenience we use, instead of $\Delta q^i$, the flavor singlet and
non-singlet combinations defined as follows:
\bea
     \Delta q^{\gamma}_S &\equiv& \sum_i~ \Delta q^i  \\
   \Delta q^i_{NS} &\equiv& \Delta q^i - \frac{\Delta q^{\gamma}_S}{n_f}
\eea
so that $\sum_i \Delta q^i_{NS} =0$ and $n_f$ is the number of relevant
active quark
flavors.  The quark-quark splitting function $\Delta {\widetilde P}_{q^i q^j}$
is made up of two pieces, the one representing the case that
$j$-quark splits into $i$-quark without through gluon, and the other one
through gluon, and may be expressed as
\be
  \Delta {\widetilde P}_{q^i q^j}=\delta_{ij}\Delta {\widetilde P}_{qq}
 + \frac{1}{n_f}~\Delta {\widetilde P}^S_{qq}  \label{SplitQQ}
\ee
where the second term is representing the splitting through gluon, and
$\Delta {\widetilde P}_{qq}$ and $\Delta {\widetilde P}^S_{qq}$ are both
independent of quark flavor, $i$ and $j$. It is noted that by construction
$\Delta {\widetilde P}^S_{qq}$ is relevant for the evolution of flavor-siglet
$\Delta q^{\gamma}_S$ and first appears in the order of $\alpha_s^2$.

In the QCD improved parton model, which is based on the factorization theorem,
the polarized virtual photon structure function $g_1^{\gamma}(x, Q^2, P^2)$
is expressed as
\bea
   g_1^{\gamma}(x, Q^2,P^2)
  &=&\int^1_x \frac{dy}{y}~\Bigl\{ \sum_i~
         \Delta q^i (y,Q^2, P^2)~C^i (\frac{x}{y}, Q^2) \nonumber  \\
 & & \qquad \qquad  +  \Delta
G^{\gamma}(y,Q^2, P^2)~C^G (\frac{x}{y}, Q^2) \Bigr\}    + C^{\gamma} (x, Q^2),
\eea
where $C^i$, $C^G$, and $C^{\gamma}$ are the coefficient functions of
$i$-quark, gluon, and photon, respectively, and are independent of
target photon mass $P^2$.  Up to one-loop level they are given by,
\bea
     C^i(z,Q^2)&=&e^2_i \Bigl\{ \delta (1-z) + \frac{\alpha_s (Q^2)}{4\pi}
      B_q(z) \Bigr\}  \label{Bq} \\
   C^G(z,Q^2)&=&<e^2> \Bigl\{ 0 + \frac{\alpha_s (Q^2)}{4\pi}
           B_G(z) \Bigr\} \label{BG} \\
   C^{\gamma}(z,Q^2)&=&\frac{\alpha}{4\pi}3 n_f <e^4> B_{\gamma}(z)
         \label{Bgamma}
\eea
where $<e^2>=\sum_i e^2_i/n_f$ and $<e^4>=\sum_i e^4_i/n_f$.
It is noted that $B_q(z)$ in Eq.(\ref{Bq}) is independent of the quark
flavor $i$.
Since $\sum_i\Delta q^i~C^i$ is rewritten as
\bea
   \sum_i \Delta q^i~C^i &=& \sum_i\Bigl\{ \Delta q^i_{NS}
  + \frac{\Delta q^{\gamma}_S}{n_f}\Bigr\}  C^i  \nonumber \\
&=& \Delta q^{\gamma}_S  (y,Q^2,P^2) <e^2>\biggl\{ \delta (1-\frac{x}{y}) +
\frac{\alpha_s (Q^2)}{4\pi}
      B_q(\frac{x}{y}) \biggr\}\nonumber \\
  & &+  \sum_i  e^2_i~ \Delta q^i_{NS} (y,Q^2,P^2)
 \biggl\{ \delta (1-\frac{x}{y}) + \frac{\alpha_s (Q^2)}{4\pi}
      B_q(\frac{x}{y}) \biggr\},
\eea
we obtain
\bea
   g_1^{\gamma}(x, Q^2,P^2)&=& \int^1_x \frac{dy}{y}~\biggl\{
                 \Delta q^{\gamma}_S (y,Q^2,P^2)~ C^S (\frac{x}{y}, Q^2) +
\Delta G^{\gamma} (y,Q^2,P^2)~C^G(\frac{x}{y}, Q^2)  \nonumber  \\
  & &  \quad + \Delta q^{\gamma}_{NS}(y,Q^2,P^2)~C^{NS} (\frac{x}{y}, Q^2)
\biggr\}   + C^{\gamma} (x, Q^2) \label{Solg}
\eea
where we have defined
\bea
  C^S(z,Q^2) &\equiv& <e^2>\biggl\{ \delta (1-z) + \frac{\alpha_s (Q^2)}{4\pi}
      B_S(z) \biggr\}  \\
  C^{NS}(z,Q^2) &\equiv&  \delta (1-z) + \frac{\alpha_s (Q^2)}{4\pi}
B_{NS}(z) \\
  \Delta q^{\gamma}_{NS}   &\equiv& \sum_i  e^2_i \Delta q^i_{NS}
    \label{DelQNS}
\eea
and $B_S(z)=B_{NS}(z)=B_q(z)$. From Eqs.(\ref{DelQi})--(\ref{SplitQQ}) and
(\ref{DelQNS}), the evolution equations for
$\Delta q^{\gamma}_S$, $\Delta G^{\gamma}$, and $\Delta q^{\gamma}_{NS}$
are now given by
\bea
  \frac{d \Delta q^{\gamma}_S(x,Q^2,P^2)}{d {\rm ln} Q^2}&=&
\int^1_x \frac{dy}{y}\biggl\{ \Bigl[
    \Delta {\widetilde P}_{q q} (\frac{x}{y}, Q^2) +
    \Delta {\widetilde P}^S_{q q} (\frac{x}{y}, Q^2) \Bigr]
  \Delta q^{\gamma}_S (y,Q^2,P^2) \nonumber  \\
     &+& n_f\Delta{\widetilde P}_{q G} (\frac{x}{y}, Q^2)~
\Delta G^{\gamma} (y,Q^2,P^2)  \biggr\}
         + \sum_i \Delta {\widetilde P}_{q^i \gamma} (x, Q^2),  \label{DeltaQS}
\nonumber \\
\\
& & \nonumber  \\
 \frac{d \Delta G^{\gamma}(x,Q^2,P^2)}{d {\rm ln} Q^2}&=& \int^1_x \frac{dy}{y}
\biggl\{ \Delta {\widetilde P}_{G q} (\frac{x}{y}, Q^2) \Delta q^{\gamma}_S
        (y,Q^2,P^2)   \nonumber  \\
  &+& \Delta {\widetilde P}_{G G} (\frac{x}{y}, Q^2)~ \Delta G^{\gamma}
(y,Q^2,P^2)  \biggr\}
  + \Delta {\widetilde P}_{G \gamma} (x, Q^2),  \\
& & \nonumber  \\
    \frac{d \Delta q^{\gamma}_{NS} (x,Q^2,P^2)}{d {\rm ln} Q^2}
 &=& \int^1_x \frac{dy}{y}
  \Delta {\widetilde P}_{q q} (\frac{x}{y}, Q^2)~ \Delta q^{\gamma}_{NS}
(y,Q^2,P^2)
    \nonumber  \\
     & & \quad  + \sum_i e^2_i \Bigl\{ \Delta {\widetilde P}_{q^i \gamma}
(x, Q^2)
             -\frac{1}{n_f} \sum_j \Delta {\widetilde P}_{q^j \gamma} (x,
Q^2) \Bigr\}~.
     \label{DeltaQNS}
\eea
Introducing a row vector $\Delta\qv^\gamma=(\Delta q^\gamma_S,~\Delta
G^\gamma,~
\Delta q^\gamma_{NS})$,
the above evolution equations, Eqs.(\ref{DeltaQS})--(\ref{DeltaQNS})
are expressed in a compact matrix form.
\be
\frac{d~\Delta\qv^\gamma(x,Q^2,P^2)}{d~ \ln Q^2}=
\Delta\kv(x,Q^2) +
\int_x^1\frac{dy}{y}\Delta\qv^\gamma(y,Q^2,P^2)\Delta P(\frac{x}{y},Q^2)
\label{evol}
\ee
where the elements of a row vector $\Delta\kv= (\Delta K_S,~\Delta
K_G,~ \Delta K_{NS})$ are
\bea
   \Delta K_S &\equiv& \sum_i \Delta {\widetilde P}_{q^i \gamma}~ ,
 \quad  \Delta K_G \equiv\Delta {\widetilde P}_{G \gamma}
   \nonumber  \\
\Delta K_{NS} &\equiv& \sum_i e^2_i \Bigl\{ \Delta {\widetilde P}_
     {q^i \gamma}
     -\frac{1}{n_f} \sum_j \Delta {\widetilde P}_{q^j \gamma}
          \Bigr\}     \label{SplitK}
\eea
Since $\Delta {\widetilde P}_{q^i \gamma}$ is proportional to $e_i^2$~, it
is easily
seen that $\Delta K_S$ and $\Delta K_{NS}$ have factors
$n_f<e^2>$ and $n_f(<e^4>-<e^2>^2)$, respectively.
The $3\times 3$ matrix $\Delta P(z,Q^2)$ is written as
\bea
\Delta P(z,Q^2)=
\left(\matrix{\Delta P^S_{q q}(z,Q^2)&\Delta P_{G q}(z,Q^2)&0\cr
           \Delta P_{q G}(z,Q^2)&\Delta P_{G G}(z,Q^2)&0 \cr
    0&0&\Delta P^{NS}_{q q}(z,Q^2)\cr}\right)
\eea
where
\bea
\Delta P^S_{q q} &\equiv& \Delta {\widetilde P}_{q q}  +
    \Delta {\widetilde P}^S_{q q}~, \quad
   \Delta P_{q G} \equiv f \Delta {\widetilde P}_{q G}  \nonumber  \\
   \Delta P_{G q} &\equiv&\Delta {\widetilde P}_{G q}~, \quad
   \Delta P_{G G} \equiv\Delta {\widetilde P}_{G G}~, \quad
   \Delta P^{NS}_{q q} \equiv \Delta {\widetilde P}_{q q}   \label{SplitP}
\eea

Once we get the information on the coefficient functions in
Eqs.(\ref{Bq})--(\ref{Bgamma}) and
parton splitting functions in Eqs.(\ref{SplitK})--(\ref{SplitP}), we can
predict the behavior of $g_1^{\gamma}(x, Q^2,P^2)$ in QCD. The NLO analysis is
now possible since the spin-dependent one-loop coefficient functions
and two-loop parton splitting functions are available~\cite{MvN,V}.
There are two methods to obtain $g_1^{\gamma}(x, Q^2,P^2)$ in NLO.
In one method, we use the parton splitting functions up to two-loop level and
we solve numerically $\Delta\qv^\gamma(x,Q^2,P^2)$ in Eq.(\ref{evol})
by iteration, starting
from the initial quark and gluon  distributions of the virtual photon at
$Q^2=P^2$. The interesting point of studying the virtual photon with mass
$-P^2$ is that  when $P^2 \gg \Lambda ^2$, the initial parton distributions
of the photon are completely known up to the one-loop level in QCD.
Then inserting the solved $\Delta\qv^\gamma(x,Q^2,P^2)$ into Eq.(\ref{Solg}),
and together with the known one-loop coefficient functions we can predict
$g_1^{\gamma}(x, Q^2,P^2)$ in NLO.

The other method, which is more common than the former, is by making use of
the inverse Mellin transformation. From now on we follow the latter method.
First we take the Mellin moments of Eq.(\ref{Solg}),
\be
     \int_0^1 dx x^{n-1}  g_1^{\gamma}(x,Q^2,P^2)
      = \Delta\qv^\gamma(n,Q^2,P^2) \cdot \Cv(n, Q^2) + C^{\gamma}(n,
Q^2)~,
\label{Gmomento}
\ee
where we have defined the moments of an arbitrary function $f(x)$ as
\be
     f(n) \equiv \int_0^1 dx x^{n-1} f(x).
\ee
Comparing Eq.(\ref{Gmomento}) with Eqs.(3.10) and (3.15),  we can
easily see the correspondence between the quantities in the QCD
improved parton model
and those in the framework of OPE as follows:
\bea
   \Bigl[ \Delta\qv^\gamma(n,Q^2,P^2)
\Bigr]_i&=& \hspace{-0.5cm} \sum_{j=S,G,NS,\gamma}
   \langle\gamma(p)|R_n^j(\mu^2=P^2)|\gamma(p)\rangle
\left(T\exp\left[\int_{{\bar g}(Q^2)}^{{\bar g}(P^2)}dg
\frac{\gamma_n(g)}{\beta(g)}
\right]\right)_{ji}  \nonumber  \\
  & &\qquad \qquad \qquad \qquad \qquad \qquad
   \qquad \qquad  (i= S,G,NS) \\
 & &  \Cv(n, Q^2)=\Cv_n(1,{\bar g}) \\
 & &  C^{\gamma}(n, Q^2)=C^{\gamma}_n(1,{\bar g}, \alpha)
\eea

Henceforce we omit the obvious
$n$-dependence for simplicity.
We expand the splitting functions $\Delta\kv(Q^2)$ and $\Delta P(Q^2)$
in powers of the QCD and QED coupling constants as
\bea
\Delta\kv(Q^2)&=&\frac{\alpha}{2\pi}\Delta\kv^{(0)}+
\frac{\alpha\alpha_s(Q^2)}{(2\pi)^2}\Delta\kv^{(1)} +\cdots \\
\Delta P(Q^2)&=&\frac{\alpha_s(Q^2)}{2\pi}\Delta P^{(0)}
+\left[\frac{\alpha_s(Q^2)}{2\pi}\right]^2 \Delta P^{(1)}
+\cdots  ~,
\eea
and introduce $t$ instead of $Q^2$ as the evolution variable~\cite{FP1},
\be
t \equiv \frac{2}{\beta_0}\ln\frac{\alpha_s(P^2)}{\alpha_s(Q^2)}~.
\ee
Then, taking the Mellin moments of the both sides in Eq.(\ref{evol}), we find
that ${\Delta\qv}^{\gamma}(t)(=\Delta\qv^\gamma(n,Q^2,P^2))$
satisfies the following inhomogenious differential equation \cite{GR,FP2}:
\bea
  \frac{d {\Delta\qv}^{\gamma}(t)}{d t}
 &=& \frac{\alpha}{2\pi} \Biggl\{ \frac{2\pi}{\alpha_s}{\Delta\kv}^{(0)}+
   \Bigl[ {\Delta\kv}^{(1)} - \frac{\beta_1}{2\beta_0}{\Delta\kv}^{(0)} \Bigr] +
   {\cal O}(\alpha_s) \Biggr\}  \nonumber \\
& &+ {\Delta\qv}^{\gamma}(t) \Biggl\{ \Delta P^{(0)}+ \frac{\alpha_s}{2\pi}
     \Bigl[ \Delta P^{(1)} - \frac{\beta_1}{2\beta_0}\Delta P^{(0)} \Bigr]  +
              {\cal O}(\alpha_s^2) \Biggr\}
\label{APE2}
\eea
where we have used the fact the QCD effective coupling constant
$\alpha_s(Q^2)$ satisfies
\be
  \frac{d \alpha_s(Q^2)}{d {\rm ln} Q^2}= -\beta_0 \frac{\alpha_s(Q^2)^2}{4\pi}
   -\beta_1 \frac{\alpha_s(Q^2)^3}{(4\pi)^2}+ \cdots
\label{beta}
\ee
with $\beta_0=11-2n_f/3$ and $\beta_1=102-38n_f/3$.
Note that the $P^2$ dependence of  $\Delta\qv^\gamma$ solely comes from the
initial condition (or boundary condition) as we will see below.

We look for the solution $\Delta\qv^\gamma(t)$ in the following form:
\be
\Delta\qv^\gamma(t)=\Delta\qv^{\gamma(0)}(t)+\Delta\qv^{\gamma(1)}(t)
\ee
where the first (second) term represents the solution in the LO (NLO).
First we discuss about the initial conditions of $\Delta\qv^\gamma$.

In Section 3, we have observed that for $ -p^2= P^2 \gg \Lambda^2$
the photon matrix elements of the hadronic operators  $R_n^i$
($i=\psi(S),G,NS$)
can be calculated perturbatively.
Choosing the square of the renormalization point $\mu^2$ to be close to
$P^2$, we obtained, to the lowest order
\be
   \langle \gamma (p) \mid R_n^i (\mu) \mid \gamma (p) \rangle =
\frac{\alpha}{4\pi}
   \Bigl( -\frac{1}{2}K_{n}^{0,i}~ {\rm ln}\frac{P^2}{\mu^2}+ A_{n}^i
\Bigr),
  \qquad  i=\psi(S),G,NS~.
\ee
The $K_{n}^{0,i}$-terms and
$A_{n}^i$-terms represent the operator mixing between the
hadronic operators and photon operators in the LO and NLO, respectively.
The operator mixing implies that there exists quark distribution in the
photon.   When we renormalize the photon matrix elements of the hadronic
operators at $\mu^2 = P^2 $, we obtain
\be
  \langle \gamma (p) \mid R_n^i (\mu) \mid \gamma (p) \rangle \vert_{\mu^2= P^2}
=
     \frac{\alpha}{4\pi}  A_{n}^i
\label{Initial}
\ee
which shows that, at $\mu^2= P^2$, quark distribution exists in the photon, not
in the LO  but in the NLO.  Thus we have
\be
  {\Delta\qv}^{\gamma(0)}(0)=0, \qquad \quad
  {\Delta\qv}^{\gamma(1)}(0)= \frac{\alpha}{4\pi}  {\Av}_n~.
\ee
Explicit expressions of ${\Av}_n$ in the
$\overline{\rm MS}$ scheme are given in Section 3.

With these initial conditions, we obtain for the solution
${\Delta\qv}^{\gamma}(t)$ of Eq.(\ref{APE2}),
\bea
{\Delta\qv}^{\gamma(0)}(t)&=&\frac{4\pi}{\alpha_s(t)}~ {\bf a}~
     \Biggl\{1-\biggl[\frac{\alpha_s(t)}{\alpha_s(0)}
   \biggr]^{1-\frac{2\Delta P^{(0)}}{\beta_0}}  \Biggr\} \label{SolLO} \\
 \nonumber  \\
{\Delta\qv}^{\gamma(1)}(t)&=& -2~{\bf a}~
  \Biggl\{  \int^t_0 d\tau  e^{(\Delta P^{(0)}-\frac{\beta_0}{2})\tau}~
\Bigl[\Delta P^{(1)}-\frac{\beta_1}{2\beta_0} \Delta P^{(0)} \Bigr]~
              e^{-\Delta P^{(0)}\tau }
  \Biggr\}~ e^{\Delta P^{(0)}t}   \nonumber  \\
& &+ {\bf b}~\Biggl\{1-\biggl[\frac{\alpha_s(t)}{\alpha_s(0)}
   \biggr]^{-\frac{2\Delta P^{(0)}}{\beta_0}} \Biggr\}
+{\Delta\qv}^{\gamma(1)}(0)\biggl[\frac{\alpha_s(t)}{\alpha_s(0)}
   \biggr]^{-\frac{2\Delta P^{(0)}}{\beta_0}}  \label{SolNLO}
\eea
where
\bea
  {\bf a}&=&\frac{\alpha}{2\pi\beta_0} {\Delta \kv}^{(0)}
\frac{1}{1-\frac{2\Delta P^{(0)}}{\beta_0}}  \\
{\bf b}&= &\Biggl\{ \frac{\alpha}{2\pi} \Bigl[ {\Delta \kv}^{(1)}
          -\frac{\beta_1}{2\beta_0} {\Delta \kv}^{(0)}
\Bigr] +2~{\bf a}~\Bigl[\Delta P^{(1)}-\frac{\beta_1}{2\beta_0} \Delta
P^{(0)} \Bigr]
\Biggr\}
   \frac{-1}{\Delta P^{(0)}}
\eea
It is noted that the parton distributions ${\Delta\qv}^{\gamma}(t)$ {\it
do} depend on
the initial conditions ${\Delta\qv}^{\gamma}(0)=(\alpha/4\pi){\Av}_n$, but
we have
seen in Section 3 that the structure function $g_1^\gamma(x,Q^2,P^2)$ itself
is independent of ${\Delta\qv}^{\gamma}(0)$ in NLO in QCD.

The moments of the splitting functions are related to the anomalous dimensions
of operators as follows:
\bea
   \Delta P^{(0)}&=&-\frac{1}{4}\widehat \gamma^0_n~, \qquad \qquad
   \Delta P^{(1)}=-\frac{1}{8}\widehat \gamma^{(1)}_n  \\
   {\Delta\kv}^{(0)}&=&\frac{1}{4}{\Kv}^0_n, \qquad \qquad \ \
    {\Delta\kv}^{(1)}=\frac{1}{8}{\Kv}^1_n
\eea
The evaluation of ${\Delta\qv}^{\gamma(0)}(t)$ and ${\Delta\qv}^{\gamma(1)}(t)$
in Eqs.(\ref{SolLO}) and (\ref{SolNLO}) can be easily done by introducing
the projection operators $P^n_i$ such as
\bea
     \Delta P^{(0)}&=&-\frac{1}{4}\widehat \gamma^0_n
=-\frac{1}{4}\sum_{i=+,-,NS}
\lambda^n_i~P^n_i,     \qquad   i=+,-,NS  \\
   P^n_i~P^n_j&=&\cases{0 &$i\ne j$, \cr
                         P^n_i &$i=j$},    \qquad
  \sum_i  P^n_i ={\bf 1}
\eea
where $\lambda^n_i$ are the eigenvalues of the matrix $\widehat \gamma^0_n$.
Then the solution ${\Delta\qv}^{\gamma}(t)$ of Eq.(\ref{APE2}) in the
NLO is written as
\bea
  {\Delta\qv}^{\gamma}(t)/\Bigl[ \frac{\alpha}{8\pi \beta_0}\Bigr] &=&
   \frac{4\pi}{\alpha_s(t)}~{\Kv}^0_n~\sum_i P^n_i~
       \frac{1}{1+\frac{\lambda^n_i}{2\beta_0}}
   \Biggl\{1-\biggl[\frac{\alpha_s(t)}{\alpha_s(0)}
   \biggr]^{1+\frac{\lambda^n_i}{2\beta_0}}  \Biggr\} \nonumber  \\
&+ &\Biggl\{{\Kv}^1_n~\sum_i P^n_i~
\frac{1}{\lambda_i^n/2\beta_0} + \frac{\beta_1}{\beta_0}{\Kv}^0_n
~\sum_i P^n_i~\Bigl( 1 - \frac{1}{\lambda_i^n/2\beta_0} \Bigr)  \nonumber  \\
&- & {\Kv}^0_n \sum_{j,i} \frac{P^n_j~ \widehat \gamma^{(1)}_n~P^n_i}
  {2\beta_0+\lambda^n_j-\lambda^n_i}~\frac{1}{\lambda^n_i/2\beta_0}~
   - 2\beta_0{\Av_n}~\sum_i P^n_i  \Biggr\} \nonumber  \\
& & \qquad \qquad \qquad \qquad \qquad \qquad \qquad \times
\Biggl\{1-\biggl[\frac{\alpha_s(t)}{\alpha_s(0)}
   \biggr]^{\frac{\lambda^n_i}{2\beta_0}}  \Biggr\}\nonumber  \\
&+&  \Biggl\{{\Kv}^0_n\sum_{i,j}~\frac{ P^n_i~\widehat \gamma^{(1)}_n
~P^n_j}{2\beta_0+\lambda^n_i-\lambda_j^n}~
\frac{1}{1+\frac{\lambda^n_i}{2\beta_0}}
 - \frac{\beta_1}{\beta_0}{\Kv}^0_n~
\sum_i~P^n_i~\frac{\lambda^n_i/2\beta_0}{1+\frac{\lambda^n_i} {2\beta_0}}
\Biggr\} \nonumber  \\
& & \qquad \qquad \qquad \qquad \qquad \qquad \qquad \times
\Biggl\{1-\biggl[\frac{\alpha_s(t)}{\alpha_s(0)}
   \biggr]^{1+\frac{\lambda^n_i}{2\beta_0}}  \Biggr\} \nonumber  \\
  &+& 2\beta_0{\Av}_n
\eea

Now inserting the above solution of ${\Delta\qv}^{\gamma}(t)$ into
Eq.(\ref{Gmomento}) and together with the information on the
coefficient functions in Eq. (3.15), we reproduce the same
formula for the moments of
$g_1^\gamma(x,Q^2,P^2)$ given in Eqs.(3.16)-(3.20), as for the case of OPE 
approach in the NLO.

\section{Sum rule for the first moment of $g_1^\gamma(x,Q^2,P^2)$}
\smallskip

The polarized structure function $g_1^\gamma$ of the real photon
satisfies a remarkable sum rule \cite{BASS,ET,NSV,FS,BBS}
\be
\int_0^1g_1^\gamma(x,Q^2)dx=0~.
\ee
Now we can ask what happens to the first moment of the virtual photon
structure function $g_1^\gamma(x,Q^2,P^2)$.
This can be studied by taking the $n \rightarrow 1$
limit of (\ref{master1}).
Note that we have the following eigenvalues of the one-loop anomalous
dimension matrix $\widehat{\gamma}^0_{n=1}$:
\begin{equation}
\lambda_{+}^{n=1}=0, \quad \lambda_{-}^{n=1}=-2\beta_0, \quad
\lambda_{NS}^{n=1}=0~.
\label{zero1}
\end{equation}

Physically speaking, the zero eigenvalues 
$\lambda_{+}^{n=1}=\lambda_{NS}^{n=1}=0$ correspond to the conservation 
of the axialvector current at one-loop order, which has the counterpart for 
the unpolarized structure function $F_2$; the conservation of energy momentum 
tensor, $\lambda_{-}^{n=2}=0$.
The other eigenvalue
$\lambda_{-}^{n=1}=-2\beta_0$, which is negative, is rather an artifact
of continuation of the anomalous dimension of the gluon operators to $n=1$, 
since there is no twist-2 gluon operator exists for $n=1$, in the RG
scheme in which only gauge-invariant operators are allowed. But $n=1$
gluon operator exists in the so called Adler-Bardeen scheme \cite{AR,BFR}.
In fact, in the QCD improved parton model approach, there is no reason why
the $n=1$ moment of the polarized gluon distribution should not be
considered \cite{AL}.

In the coefficients
\begin{equation}
\tilde{P}_i^n\frac{1}{1+\lambda_i^n/2\beta_0},\quad
{\cal A}_i^n, \quad {\cal B}_i^n \qquad
(i=+,-,NS)
\end{equation}
the special points (\ref{zero1}) would develop the singularities
at $n=1$, since in those coefficients there exist the factors
\begin{equation}
\frac{1}{\lambda_{+}^{n}}, \quad \frac{1}{\lambda_{NS}^{n}},
\quad \frac{1}{1+\lambda_{-}^{n}/2\beta_0}.
\end{equation}
Now if we take the limit of $n$ going to 1, we have
\begin{eqnarray}
&&\tilde{P}_i^n\frac{1}{1+\lambda_i^n/2\beta_0} \rightarrow 0 \qquad
(i=+,-,NS) \nonumber\\
&&{\cal A}_{+}^n \rightarrow \mbox{finite}, \quad
{\cal A}_{-}^n \rightarrow 0, 
\quad {\cal A}_{NS}^n \rightarrow \mbox{finite} \nonumber\\
&&{\cal B}_{+}^n \rightarrow 0, 
\quad {\cal B}_{-}^n \rightarrow \mbox{finite},
\quad {\cal B}_{NS}^n \rightarrow 0~.
\end{eqnarray}
However, ${\cal A}_{+}^n $, ${\cal A}_{NS}^n$, and ${\cal B}_{-}^n$
are multiplied by the following vanishing factors
\begin{equation}
\left\{1-\left(\frac{\alpha_s(Q^2)}
{\alpha_s(P^2)}\right)^{\lambda_{+}^n/2\beta_0}\right\}, \quad
\left\{1-\left(\frac{\alpha_s(Q^2)}
{\alpha_s(P^2)}\right)^{\lambda_{NS}^n/2\beta_0}\right\}, \quad
\left\{1-\left(\frac{\alpha_s(Q^2)}{\alpha_s(P^2)}\right)
^{\lambda_{-}^n/2\beta_0+1}\right\}~,  \label{vanishing}
\end{equation}
respectively, and thus the terms proportional to ${\widetilde P}^n_i$,
${\cal A}_i^n$, and ${\cal B}_i^n$ in Eq.(\ref{master1})
all vanish in the $n=1$ limit. Note that these vanishing factors are specific
to the case of the virtual photon target, and that such factors do not appear
when the target is real photon.

Thus the first three terms in the 1st moment vanish
irrespective of the RG scheme.
So we get
\begin{equation}
\int_0^1dx g_1^\gamma(x,Q^2,P^2)=\frac{\alpha}{4\pi}\frac{1}{2\beta_0}
{\cal C}^{n=1} +{\cal O}(\alpha_s)
\end{equation}

Now let us consider ${\cal C}^{n=1}$, which is given by
\be
{\cal C}^{n=1}=12\beta_0<e^4>(B_G^n+{\tilde A}_{nG}^\psi)|_{n=1}
\ee
As we have seen, the combination $(B_G^n+\widetilde A_{nG}^\psi)$ is
renormalization-scheme independent~\cite{BBDM}.  The results in the
$\overline{\rm MS}$ scheme~\cite{MvN,V,BQ} are
\be
B_G^{n=1}=0, \qquad {\tilde A}_{n=1\ G}^\psi=-2n_f
\ee
The same results have been obtained by
Kodaira \cite{JK} in the framework of OPE and RG method. He set $B_G^{n=1}=0$,
observing that there is no gauge-invariant twist-2 gluon operator for $n=1$ and
obtained ${\tilde A}_{n=1\ G}^\psi=-2n_f$ from the Adler-Bell-Jackiw anomaly.
In the end, we have for the sum rule of the virtual photon
structure function $g_1^\gamma$,
\be
\int_0^1 dx g_1^\gamma(x,Q^2,P^2)=-\frac{3\alpha}{\pi}\sum_{i=1}^{n_f}
e_i^4 +{\cal O}(\alpha_s),  
\ee
where
\be
Q^2 \gg P^2 \gg m_i^2, \Lambda^2, \quad i=1,\cdots, n_f
\ee
with $m_i$, the mass of $i$-th flavor quark, and $n_f$, the number of 
active flavors.

Now it should be pointed out that we can further pursue the QCD
corrections of order $\alpha_s$ to the first moment of $g_1^\gamma$.
In the above equation for the first moment, the leading order is ${\cal O}(1)$
not of order $1/\alpha_s(Q^2)$, which is the case for the general moments. So
we now go to the order $\alpha_s$ QCD correction.

First we take the renormalization scheme of Kodaira \cite{JK}. 
We write down the first moment of $g_1^\gamma(x,Q^2,P^2)$:
\bea
\int_0^1 dx g_1^\gamma(x,Q^2,P^2)
&=& C_1^\psi(Q^2/P^2,{\bar g}(P^2),\alpha)\langle \gamma(p)|
R_1^\psi(\mu^2=P^2)|\gamma(p)\rangle \nonumber\\
&+&
 C_1^{NS}(Q^2/P^2,{\bar g}(P^2),\alpha)\langle \gamma(p)|
R_1^{NS}(\mu^2=P^2)|\gamma(p)\rangle \quad
\eea
Here it should be emphasized that because of the absence of the
gauge-invariant $n=1$ gluon and photon operators, the mixing problem 
becomes much simpler. The coefficient functions can be given by
\be
\pmatrix{C_1^\psi(Q^2/P^2,{\bar g}(P^2),\alpha)\cr
C_1^{NS}(Q^2/P^2,{\bar g}(P^2),\alpha)}
=T \exp\int_{{\bar g}(Q^2)}^{{\bar
g}(P^2)}dg'\frac{{\widehat\gamma}_1(g')}{\beta(g')}
\pmatrix{C_1^\psi(1,{\bar g}(Q^2),\alpha)\cr
C_1^{NS}(1,{\bar g}(Q^2),\alpha)}
\ee
where ${\widehat\gamma}_1(g)$ is a $2\times 2$ diagonal matrix:
\be
{\widehat\gamma}_1(g)=
\left(
\begin{array}{c|c}
\gamma_{\psi\psi}(g)&0\\
\hline
0&\gamma_{NS}(g)
\end{array}
\right)~.
\ee
Here anomalous dimensions are expanded in powers of the coupling constant:
\be
\gamma(g)=\gamma^{(0)}\frac{g^2}{16\pi^2}+\gamma^{(1)}(\frac{g^2}{16\pi^2})^
2+{\cal O}(g^6)~,
\ee
where
\be
\gamma_{\psi\psi}^{(0)}=\gamma_{NS}^{(0)}=0,\quad
\gamma_{\psi\psi}^{(1)}=24C_F
T_f=24\cdot\frac{N_c^2-1}{2N_c}\cdot\frac{n_f}{2}=16n_f,\quad
\gamma_{NS}^{(1)}=0~,
\ee
and the coefficient functions are
\bea
C_1^\psi(1,{\bar
g}(Q^2),\alpha)&=&<e^2>(1-\frac{3}{4}C_F\frac{\alpha_s(Q^2)}{\pi})
=<e^2>(1-\frac{\alpha_s(Q^2)}{\pi})\nonumber\\
C_1^{NS}(1,{\bar g}(Q^2),\alpha)&=&
1\cdot (1-\frac{3}{4}C_F\frac{\alpha_s(Q^2)}{\pi})
=1-\frac{\alpha_s(Q^2)}{\pi}
\eea
\be
T \exp\int_{{\bar g}(Q^2)}^{{\bar
g}(P^2)}dg'\frac{{\widehat\gamma}_1(g')}{\beta(g')}
={\bf 1}-\frac{1}{16\pi^2}\frac{{\widehat\gamma}^{(1)}}{2\beta_0}
[{\bar g}^2(P^2)-{\bar g}^2(Q^2)]~.
\ee

Here we have the finite matrix element of the quark operators
between the virtual photon states:
\be
\langle \gamma(p)|R_{n=1}^i(\mu^2=P^2)|\gamma(p)\rangle=
\frac{\alpha}{4\pi}A_{n=1}^i,
\ee
where
\be
{\Av}_n=6(<e^2>,0,<e^4>-<e^2>^2){\tilde A}^\psi_{nG}.
\ee
Now we recall Kodaira's statement that the bare Green's function for the
$n=1$ case does not receive divergent corrections but the finite correction
connected with Adler-Bell-Jackiw anomaly:
\be
{\tilde A}^\psi_{n=1\ G}=-4T(R)=-2n_f.
\ee

Putting all these equations together, we finally obtain the
${\cal O}(\alpha_s)$ QCD correction:
\bea
\int_0^1dx g_1^\gamma(x,Q^2,P^2)
&=&-\frac{3\alpha}{\pi}
\left[\sum_{i=1}^{n_f}e_i^4\left(1-\frac{\alpha_s(Q^2)}{\pi}\right)
\right.\nonumber\\
&-&\left.\frac{2}{\beta_0}(\sum_{i=1}^{n_f}e_i^2)^2\left(
\frac{\alpha_s(P^2)}{\pi}-\frac{\alpha_s(Q^2)}{\pi}\right)\right]
+{\cal O}(\alpha_s^2). \label{Oalpha}
\eea
This result is perfectly in agreement with the one obtained by
Narison, Shore and Veneziano in ref.\cite{NSV}, apart from the
overall sign for the definition of $g_1^\gamma$.

Now we show that we obtain the same result for the first moment of
$g_1^\gamma(x,Q^2,P^2)$ in the $\overline{\rm MS}$ scheme. Although
there exist no gauge-invariant twist-2 gluon- and photon-operators for
$n=1$, the $\overline{\rm MS}$ calculation of the anomalous
dimensions gives non-zero results for $\gamma_{GG}^{n=1}$ and
$\gamma_{G\psi}^{n=1}$. Thus the $\overline{\rm MS}$-scheme results
rather correspond to the QCD parton model approach where
the first moments of the gluon and photon distributions are as well
defined as the other distributions.

Including the gluon and photon operators, let us start with Eq.(3.13)
for the first moment of $g_1^\gamma(x,Q^2,P^2)$ in $\overline{\rm MS}$
scheme:
\bea
\int_0^1 dx g_1^\gamma(x,Q^2,P^2)
&=&\frac{\alpha}{4\pi}\Av_1\cdot M_1(Q^2/P^2,{\bar g}(P^2)) \cdot
\Cv_1(1,{\bar g}(Q^2)) \nonumber\\
&+&\Xv_1(Q^2/P^2,{\bar g}(P^2),\alpha) \cdot \Cv_1(1,{\bar g}(Q^2))
+C_1^\gamma \label{firstMom}
\eea
We expand $\Av_1$ in powers of ${\bar g}^2(P^2)$ as
\be
   \Av_1=\Av_1^{(0)}+ \frac{{\bar g}^2(P^2)}{16\pi^2}\Av_1^{(1)} + \cdots
\label{ExpandA}
\ee
where $\Av_1^{(0)}$ is given in Eq.(3.23).
Using the $\overline{\rm MS}$ results for the $n=1$ anomalous dimensions,
we obtain up to the order  ${\cal O}(g^2)$
 \bea
 M_1
 &=&
 \left(
 \begin{array}{ccc}
 1-\frac{1}{2\beta_0}\frac{1}{16\pi^2}
 \left[{\bar g}^2(P^2)-{\bar g}^2(Q^2)\right]\times 24C_FT_f,
 &\cdots, &0 \\
 0,&\cdots, &0 \\
 0,&\cdots,&1
 \end{array}
 \right)
 \eea
where the second column is irrelevant since
$B_G^1=0$ in $\overline{\rm MS}$ scheme and thus $C_1^G(1,{\bar g})$
starts in ${\cal O}(\alpha_s^2)$.
Now it is easy to see that the first term of Eq.(\ref{firstMom}),
to be more specific, $(\frac{\alpha}{4\pi} \Av_1^{(0)} M_1
\Cv_1(1,{\bar g}(Q^2)))$ gives the same result as in Eq.(\ref{Oalpha}).

Let us now consider the contributions of other terms.
If $A_1^{\psi(1)}$ and $A_1^{NS(1)}$ in the second term in
Eq.(\ref{ExpandA}) remain non-zero, then they give the ${\cal O}(g^2)$
contribution. But $A_1^{\psi(1)}=A_1^{NS(1)}=0$ due to the
non-renormalization theorem \cite{AB} for the triangle anomaly,
so its contribution is at most in  ${\cal O}(\alpha_s^2)$.
The contribution of the second term in Eq.(\ref{firstMom}) is also
in ${\cal O}(\alpha_s^2)$, since $\Kv_1^0=\Kv_1^{(1)}=0$ and we expect
\be
      K_{\psi}^{(2),n=1}=K_{NS}^{(2),n=1}=0~,
\ee
for the three-loop mixing anomalous dimensions
which are implied from the fact that the three-loop
$\gamma_{\psi G}^{(2),n=1}=0$~\cite{AL}.

Finally we expand the third term of Eq.(\ref{firstMom}), $C_1^\gamma$,  as
\be
 C_1^{\gamma}(1,{\bar g},\alpha)=\frac{\alpha}{4\pi}\delta_\gamma
   \Bigl[B^{(0),n=1}_\gamma + \frac{{\bar g}^2}{16\pi^2}
B^{(1),n=1}_\gamma
  +\cdots  \Bigr]
\ee
where $B^{(0),n=1}_\gamma =B^{n=1}_\gamma$ in Eq.(3.15).
We already know that $B^{(0),n=1}_\gamma =0$ in $\overline{\rm MS}$
scheme. On the other hand, the two-loop (${\cal O}(\alpha_s^2)$)
coefficient function for the polarized gluon has been caluculated
in the $\overline{\rm MS}$ scheme
by Zijlstra and van Neerven~\cite{ZvN}. It is made up of
two terms, one  proportional to factor $C_F T_f n_f$ and the other
proportional to factor $C_A T_f n_f$. The first moments of both terms
turn out to vanish. It can be shown that
the two-loop (${\cal O}(\alpha \alpha_s)$)
coefficient function for the polarized photon, $B^{(1)}_\gamma$, is
obtained from  the two-loop gluon coefficient function, by picking up the
term  with  the $C_F T_f n_f$ factor and by modifying the group factors.
Thus we conclude that the first moment of $B^{(1)}_\gamma$ is zero.
In the end, $C_1^\gamma$ does not give ${\cal O}(1)$ nor
${\cal O}(\alpha_s)$
contributions  to the first moment of $g_1^\gamma(x,Q^2,P^2)$.
This means that we arrive at the same result for the 1st moment
of $g_1^\gamma$ given in (\ref{Oalpha}) in the $\overline{\rm MS}$ scheme.

\section{Numerical analysis}
\smallskip

We now perform the inverse Mellin transform of the equation
(\ref{master1}) to get $g_1^\gamma$ as a function of $x$.
The $n$-th moment is denoted as
\be
M(n,Q^2,P^2)=\int_0^1x^{n-1}g_1^\gamma(x,Q^2,P^2)dx.
\label{mom}
\ee
Then by inverting the moments (\ref{mom}) we get
\be
g_1^\gamma(x,Q^2,P^2)=\frac{1}{2\pi i}
\int_{C-i\infty}^{C+i\infty}M(n,Q^2,P^2)x^{-n}dn
\ee
In order to have better convergence of the numerical intergration,
we change the contour in the complex $n$-plane
from the vertical line connecting $C-i\infty$
with $C+i\infty$ (C is an appropriate positive constant), introducing a 
small positive constant $\varepsilon$, to
\be
n=C-\varepsilon|y|+iy, \qquad\qquad -\infty <y< \infty
\ee

Hence we have
\bea
g_1^\gamma(x,Q^2,P^2)&=&\frac{1}{2\pi i}\int_0^\infty 
M(C-\varepsilon y+iy,Q^2,P^2)
e^{-(C-\varepsilon y+iy)\log(x)}(i-\varepsilon)dy \nonumber\\
&+&\frac{1}{2\pi i}\int_{-\infty}^0 M(C+\varepsilon y+iy,Q^2,P^2)
e^{-(C+\varepsilon y+iy)\log(x)}(i+\varepsilon)dy \nonumber\\
&=&\frac{1}{\pi}\int_0^\infty
\left[\mbox{Re}\{M(z,Q^2,P^2)e^{-z\log(x)}
\} \right.\nonumber\\
&&\qquad\qquad-\left.\varepsilon\mbox{Im}\{M(z,Q^2,P^2)
e^{-z\log(x)}\}\right]dy \nonumber\\
&&z=C-\varepsilon y+iy ~. \label{invmom}
\eea

In Fig.3 we have plotted, as an illustration, the result for
$n_f=3$, $Q^2=30$ GeV$^2$ and $P^2=1$ GeV$^2$ for the QCD scale parameter
$\Lambda=0.2$ GeV. The vertical axis corresponds to
\be
g_1^\gamma(x,Q^2,P^2)/\frac{3\alpha}{\pi}n_f<e^4>\ln\frac{Q^2}{P^2}.
\label{normalized}
\ee
Here we have shown four cases; the Box (tree) diagram contribution;
\be
g_1^{\gamma({\rm Box})}(x,Q^2,P^2)
=(2x-1)\frac{3\alpha}{\pi}n_f<e^4>\ln\frac{Q^2}{P^2}~,
\ee
the Box diagram contribution including non-leading correction 
ignoring quark mass
\be
g_1^{\gamma({\rm Box,non-leading})}(x,Q^2,P^2)=\frac{3\alpha}{\pi}n_f<e^4>
\left[(2x-1)\ln\frac{Q^2}{P^2}-2(2x-1)(\ln x+1)\right]~,
\ee
the leading-order (LO) QCD correction and the next-to-leading order (NLO)
QCD correction. We observe that the NLO QCD correction is significant at large
$x$ as well as at low $x$.
We have also studied other examples with different $Q^2$ and $P^2$. In Fig.4
we have plotted the case for $Q^2=100$ GeV$^2$ with $P^2=1$ GeV$^2$. Another 
case for $Q^2=30$ GeV$^2$ with $P^2=3$ GeV$^2$ is shown in Fig.5.
We have not seen any sizable change for the normalized structure
function (\ref{normalized}) for these different values of $Q^2$ and $P^2$. 
We examined the $n_f=4$ case as well. It is observed that the normalized
structure function is insensitive to the number of active flavors. Here
we have not directly taken into account the heavy quark mass dependence,
but rather confined ourselves to the above kinematical region. 
It turns out from the numerical analyis as well as from theoretical arguments
that, as $P^2$ increases, the NLO QCD result approaches the Box
contribution including the non-leading correction, as in the unpolarized
structure function \cite{UW1,UW2}.

Now let us cosider the real photon case $P^2=0$. The structure function can be
decomposed as
\be
g_1^\gamma(x,Q^2)=g_1^\gamma(x,Q^2)\vert_{\rm pert.} +
g_1^\gamma(x,Q^2)\vert_{\rm non-pert.}
\ee
The second term can only be computed by some non-perturbative method like
lattice QCD,
or estimated by vector meson dominance model (VMD). The first term, the
point-like
piece, can be calculated in a perturbative method. Actually, it can formally be
recovered in our analysis by setting $P^2=\Lambda^2$ in Eq.(\ref{master1}).
In Fig.6, we have plotted the point-like piece of $g_1^\gamma$ of the real
photon. The LO QCD result coincides with the previous result 
obtained by Sasaki in \cite{KS}. The NLO result is qualitatively consistent 
with the analysis by Stratmann and Vogelsang \cite{SV}.
Finally, a comment on the $n=1$ limit of the real photon structure function
is in order. In the case of the unpolarized structure function $F_2^\gamma$
we have a singularity of ${\cal A}_{-}^n$ at $n=2$ due to the vanishing of
$\lambda_{-}^n$ at $n=2$ which leads to the negative structure 
function \cite{DO}. As discussed in refs.\cite{BILL,AG,DUKE} we have to
introduce some regularization prescription to recover positive structure
function. For the polarized case we do not have such complication at
$n=1$ as we have seen in sect. 5.

\section{Conclusion}
\smallskip

Here in the present paper, we have investigated virtual photon's spin
structure function $g_1^\gamma(x,Q^2,P^2)$ for the kinematical region
$\Lambda^2\ll P^2 \ll Q^2$, in the next-to-leading order in QCD.
We presented our arguments both in the framework of OPE supplemented by RG 
method and in the DGLAP equation approach.
The results are shown to be independent of the renormalization scheme.

The first moment of $g_1^\gamma$ for the virtual photon is non-vanishing
in contrast to the vanishing first moment for the real photon case.
We can go a step further to the order $\alpha_s$ which has turned out
to reproduce
the previous result of Narison, Shore and Veneziano \cite{NSV}, and
the result is RG scheme-independent.

The numerial evaluation of $g_1^\gamma$ by the inverse Mellin transform
was performed. The result shows that the NLO QCD corrections are significant
at large $x$ and also at small $x$. The numerical analysis can also be applied
to the point-like component of the real photon structure function. The result
is qualitatively consistent with the previous analysis.

Although we have neglected in our kinematical region, we should also
consider the power corrections of the form $(P^2/Q^2)^k (k=1,2,\cdots)$,
which are arising from the target mass effects as well as from
higher-twist effects.

In the present paper we only presented the result for the polarized
photon structure function $g_1^\gamma$ itself. In the course of the
parton model analysis, we also obtain the polarized parton
distributions \cite{SVZ,GSV} of the longitudinally polarized photon,
for the case of virtual photon, which will be discussed elsewhere.

As a future subject, it would be intriguing to study another spin
structure function $g_2^\gamma$
which only exists for off-shell photon. In the OPE language, the twist-2 as 
well as twist-3 operators contribute to the QCD effects for $g_2^\gamma$, 
which are now under investigation.

\vspace{5cm}

\vspace{0.5cm}
\leftline{\large\bf Acknowledgement}
\vspace{0.5cm}

We thank G. Altarelli, J. Kodaira, G. Ridolfi, G. M. Shore, 
G. Veneziano and W. Vogelsang for valuable discussions. 
Part of this work was done while one of us (T.U.) was at CERN. 
He thanks the CERN Theory Division for the kind hospitality.
This work is partially supported by the Monbusho Grant-in-Aid
for Scientific Research No.(C)(2)-09640342 and No.(C)(2)-09640345.

\newpage
\appendix

{\LARGE\bf Appendix}

\section{Notation for anomalous dimensions}

To the lowest order in $\alpha$, the anomalous dimension matrix has the form
\be
      \gamma_{n}(g,\alpha)=\pmatrix{{\widehat \gamma}_{n}(g^2)&0\cr
    {\Kv}_{n}(g^2,\alpha)&0\cr}
\ee
where ${\widehat \gamma}_{n}(g^2)$ is the usual $3\times 3$ anomalous
dimension matrix in the hadronic sector
\be
  {\widehat \gamma}_{n}(g)=\pmatrix{\gamma^n_{\psi\psi}(g)&
                              \gamma^n_{G\psi}(g) &0\cr
    \gamma^n_{\psi G}(g)&\gamma^n_{GG}(g)&0\cr
   0&0&\gamma^n_{NS}(g)\cr} ~,
\ee
and ${\Kv}_{n}(g,\alpha)$ is the three-component row vector
\be
   {\Kv}_{n}(g,\alpha)=\pmatrix{K^n_{\psi}(g,\alpha),&
    K^n_{G}(g,\alpha),&K^n_{NS}(g,\alpha)\cr}
\ee
representing the mixing between photon and three hadronic operators.

\noindent
The anomalous dimensions are expanded as
\bea
   {\widehat \gamma}_{n}(g)&=&\frac{g^2}{16\pi^2}{\widehat \gamma}^{0}_{n}
   +\frac{g^4}{(16\pi^2)^2}{\widehat \gamma}^{(1)}_{n}+{\rm O}(g^6),  \\
   {\Kv}_{n}(g,\alpha)&=&-\frac{e^2}{16\pi^2}{\Kv}^{0}_{n} -
    \frac{e^2 g^2}{(16\pi^2)^2}{\Kv}^{(1)}_{n}+{\rm O}(e^2 g^4) .
\eea

\noindent
The one-loop anomalous dimension matrix ${\widehat \gamma}^{0}_{n}$ can be
expressed in terms of its eigenvalues $\lambda^n_i~ (i=+, -, NS)$ as
\be
  {\widehat \gamma}^{0}_{n}=\sum_{i=+, -, NS} \lambda^n_i~P^n_i
\ee
where
\bea
     \lambda^n_{\pm}&=&\frac{1}{2}\Bigl\{\gamma^{0,n}_{\psi\psi}+
 \gamma^{0,n}_{GG} \pm \Bigl[(\gamma^{0,n}_{\psi\psi}-
\gamma^{0,n}_{GG})^2+4 \gamma^{0,n}_{\psi G}\gamma^{0,n}_{G\psi} \Bigr]^{1/2}
             \Bigr\}  \\
 \lambda^n_{NS}&=&\gamma^{0,n}_{NS}
\eea
and $P^n_i$ are the corresponding projection operators,
\bea
   P^n_{\pm}&=&\frac{1}{\lambda^n_{\pm}-\lambda^n_{\mp}}
   \pmatrix{\gamma^{0,n}_{\psi\psi}-\lambda^n_{\mp}&\gamma^{0,n}_{G\psi}&0\cr
 \gamma^{0,n}_{\psi G}& \gamma^{0,n}_{GG}-\lambda^n_{\mp}&0\cr
             0&0&0\cr}  \\
    P^n_{NS}&=&\pmatrix{0&0&0\cr 0&0&0\cr 0&0&1\cr}
\eea

\section{Explicit expressions for anomalous dimensions}

\subsection{One-loop order}

\bea
  \gamma^{0,n}_{\psi\psi}&=&\gamma^{0,n}_{ NS}=
    2C_F\Bigl[-3-\frac{2}{n(n+1)}+4S_1(n)  \Bigr]~, \\
  \gamma^{0,n}_{\psi G}&=&-8T_f~\frac{n-1}{n(n+1)}~,  \\
  \gamma^{0,n}_{G \psi}&=&-4C_F\frac{n+2}{n(n+1)}~,  \\
  \gamma^{0,n}_{G G}&=&2C_A\Bigl[-\frac{11}{3}-\frac{8}{n(n+1)}+4S_1(n)
         \Bigr]+\frac{8}{3}T_f~.
\eea
where
\be
S_1(n)=\sum_{j=1}^n\frac{1}{j}.
\ee
and

\be
   C_A=3, \quad C_F=\frac{4}{3}, \quad T_f=T_Rn_f=n_f/2 
\ee
with $n_f$ being the number of flavors.

\bea
   {\Kv}^{0}_{n}&=&\pmatrix{K^{0,n}_{\psi},&0,&K^{0,n}_{ NS}} ~,\\
   K^{0,n}_{\psi}&=&24~n_f<e^2>\frac{n-1}{n(n+1)}~,\\
   K^{0,n}_{ NS}&=&24~n_f(<e^4>-<e^2>^2) \frac{n-1}{n(n+1)}~.
\eea

\subsection{Two-loop order \cite{MvN,V}}

\subsubsection{Non-singlet sector}

\be
\gamma^{(1),n}_{NS}=8 C_F^2~A_{NS}^n + 8C_A C_F~ B_{NS}^n +
         8 C_F T_f~D_{NS}^n~,
\ee
with
\bea
    A_{NS}^n&=&-\frac{3}{8}+\frac{5}{n}-\frac{5}{n+1}-\frac{3}{n^2}
-\frac{2}{(n+1)^2}+ \frac{1}{n^3} -\frac{3}{(n+1)^3}  \nonumber \\
 & &+(-1)^n \Bigl\{-\frac{4}{n}+\frac{4}{n+1}+\frac{2}{n^2}
+\frac{2}{(n+1)^2}- \frac{2}{n^3} +\frac{2}{(n+1)^3}    \Bigr\} \nonumber \\
 & &+S_1(n)\Bigl(\frac{2}{n^2}-\frac{2}{(n+1)^2} \Bigr)
        +S_2(n)\Bigl(3-\frac{2}{n}+\frac{2}{n+1}+4S_1(n) \Bigr) \nonumber \\
& &+S'_2(\frac{n}{2})\Bigl(\frac{2}{n}-\frac{2}{n+1}-4S_1(n) \Bigr)
     -S'_3(\frac{n}{2})+8\widetilde S(n) \\
& &  \nonumber  \\
  B_{NS}^n&=&-\frac{17}{24}-\frac{187}{18}\frac{1}{n}+\frac{187}{18}
\frac{1}{n+1}+\frac{17}{6}\frac{1}{n^2}-\frac{5}{6}\frac{1}{(n+1)^2}
- \frac{1}{n^3} +\frac{1}{(n+1)^3}\nonumber \\
& &+(-1)^n \Bigl\{ ~\frac{2}{n}-
\frac{2}{n+1}-\frac{1}{n^2}-\frac{1}{(n+1)^2}
+ \frac{1}{n^3} -\frac{1}{(n+1)^3}    \Bigr\} \nonumber \\
& &+\frac{67}{9}S_1(n)+S_2(n)\Bigl(-\frac{11}{3}+\frac{2}{n}-
\frac{2}{n+1}-4S_1(n)\Bigr) \nonumber \\
& &+S'_2(\frac{n}{2})\Bigl(-\frac{1}{n}+\frac{1}{n+1}+2S_1(n) \Bigr)
 +\frac{1}{2} S'_3(\frac{n}{2})-4\widetilde S(n) \\
& &  \nonumber  \\
D_{NS}^n&=&\frac{1}{6}+\frac{22}{9}\frac{1}{n}-\frac{22}{9}
\frac{1}{n+1}-\frac{2}{3}\frac{1}{n^2}+\frac{2}{3}\frac{1}{(n+1)^2}
\nonumber \\
  & &-\frac{20}{9}S_1(n)+\frac{4}{3}S_2(n)
\eea

where
\be
S_2(n)=\sum_{j=1}^n \frac{1}{j^2},\quad
S_3(n)=\sum_{j=1}^n \frac{1}{j^3},\quad
\widetilde S(n)=\sum_{j=1}^n \frac{(-1)^j}{j^2}S_1(j),\quad
\ee
and
\bea
S'_2(\frac{n}{2})&=&\frac{1+(-1)^n}{2}S_2(\frac{n}{2})
+\frac{1-(-1)^n}{2}S_2(\frac{n-1}{2}) \\
S'_3(\frac{n}{2})&=&\frac{1+(-1)^n}{2}S_3(\frac{n}{2})
+\frac{1-(-1)^n}{2}S_3(\frac{n-1}{2})~.
\eea

\bigskip

\subsubsection{Singlet sector}

[1]\quad $\gamma_{\psi\psi}$:
\be
 \gamma^{(1),n}_{\psi\psi}=\gamma^{(1),n}_{NS}+\gamma^{(1),n}_{PS,\psi\psi}
\ee

with
\be
  \gamma^{(1),n}_{PS,\psi\psi}=8C_F T_f \Biggl\{-\frac{2}{n}+\frac{2}{n+1}-
\frac{2}{n^2}+\frac{6}{(n+1)^2}+ \frac{4}{n^3}+ \frac{4}{(n+1)^3}
\Biggr\}
\ee

\bigskip

\noindent
[2] \quad $\gamma_{\psi G}$:
\be
\gamma^{(1),n}_{\psi G}=8 C_F T_f~ D_{\psi G}+8 C_A T_f~ E_{\psi G}~,
\ee

with
\bea
    D_{\psi G}^n&=& S_1^2(n)\Bigl(\frac{2}{n}-\frac{4}{n+1} \Bigr)
  - S_2(n)\Bigl(\frac{2}{n}-\frac{4}{n+1} \Bigr)
  + S_1(n)\Bigl(~\frac{8}{n}-\frac{8}{n+1}-\frac{4}{n^2}   \Bigr) \nonumber \\
 & &+\frac{22}{n}-\frac{27}{n+1}-\frac{9}{n^2}-
\frac{8}{(n+1)^2}+ \frac{2}{n^3} +\frac{4}{(n+1)^3}  \\
& &  \nonumber \\
 E_{\psi G}^n&=&-\frac{24}{n}+\frac{22}{n+1}+\frac{2}{n^2}+
\frac{24}{(n+1)^2}+ \frac{4}{n^3} +\frac{24}{(n+1)^3} \nonumber \\
 & &- S_1(n)\Bigl(~\frac{8}{n}-\frac{8}{n+1}-\frac{8}{(n+1)^2}   \Bigr)
 -  S_1^2(n)\Bigl(~\frac{2}{n}-\frac{4}{n+1} \Bigr)  \nonumber \\
& &+ S_2(n)\Bigl(~\frac{2}{n}-\frac{4}{n+1} \Bigr)
  - S'_2(\frac{n}{2}) \Bigl(~\frac{2}{n}-\frac{4}{n+1} \Bigr)  \nonumber \\
& &+\Bigl[1+(-1)^n \Bigr] \Bigl(~\frac{2}{n}-\frac{4}{n+1} \Bigr)
\Bigl(-2 S_2(n) +  S'_2(\frac{n}{2}) + \zeta(2) \Bigl)
\eea

\bigskip

\noindent
[3]\quad $\gamma_{G \psi}$:

\be
\gamma^{(1),n}_{G \psi}=8 C_F^2~A_{G \psi}^n + 8 C_A C_F~B_{G \psi}^n +
         8 C_F T_f~ D_{G \psi}^n,
\ee
with
\bea
  A_{G \psi}^n&=&\frac{17}{2}\frac{1}{n}-\frac{4}{n+1}-\frac{2}{n^2}-
\frac{1}{2}\frac{1}{(n+1)^2}- \frac{2}{n^3} -\frac{1}{(n+1)^3} \nonumber \\
& &- S_1(n)\Bigl(~\frac{2}{n}+\frac{1}{n+1}+\frac{2}{(n+1)^2}   \Bigr)
 +  S_1^2(n)\Bigl(\frac{2}{n}-\frac{1}{n+1} \Bigr) \nonumber \\
& &+ S_2(n)\Bigl(\frac{2}{n}-\frac{1}{n+1} \Bigr)  \\
& &  \nonumber \\
B_{G \psi}^n&=&-\frac{41}{9}\frac{1}{n}-\frac{35}{9}\frac{1}{n+1}
+\frac{4}{n^2}-\frac{38}{3}\frac{1}{(n+1)^2}- \frac{4}{n^3}
-\frac{6}{(n+1)^3} \nonumber \\
& &+
S_1(n)\Bigl(~\frac{10}{3}\frac{1}{n}+\frac{1}{3}\frac{1}{n+1}+\frac{4}{n^2}
\Bigr)
 - S_1^2(n)\Bigl(~\frac{2}{n}-\frac{1}{n+1} \Bigr) \nonumber \\
& &- S_2(n)\Bigl(~\frac{2}{n}-\frac{1}{n+1}\Bigr) + S'_2(\frac{n}{2})
\Bigl(~\frac{2}{n}-\frac{1}{n+1}\Bigr) \nonumber \\
& &+\Bigl[1+(-1)^n \Bigr] \Bigl(~\frac{2}{n}-\frac{1}{n+1}\Bigr)
\Bigl( 2S_2(n) - S'_2(\frac{n}{2})-\zeta(2) \Bigr)
  \\
& &  \nonumber \\
 D_{G \psi}^n&=&\frac{16}{9}\frac{1}{n}+\frac{4}{9}\frac{1}{n+1}+
\frac{4}{3}\frac{1}{(n+1)^2}
- S_1(n)\Bigl(~\frac{8}{3}\frac{1}{n}-\frac{4}{3}\frac{1}{n+1}  \Bigr)
\eea

\bigskip

\noindent
[4] \quad $\gamma_{GG}$:
\be
\gamma^{(1),n}_{GG}=8C_F T_f~ D_{GG}^n+8C_A T_f~ E_{GG}^n+8 C^2_A~ F_{GG}^n~,
\ee
with
\bea
    D_{GG}^n&=&1+\frac{10}{n}-\frac{10}{n+1}-\frac{10}{n^2}+
\frac{2}{(n+1)^2}+ \frac{4}{n^3}+\frac{4}{(n+1)^3}   \\
& &   \nonumber \\
 E_{GG}^n&=&\frac{4}{3}+\frac{76}{9}\frac{1}{n}-\frac{76}{9}\frac{1}{n+1}
-\frac{4}{3}\frac{1}{n^2}-\frac{4}{3}\frac{1}{(n+1)^2}-\frac{20}{9} S_1(n)  \\
& &   \nonumber \\
 F_{GG}^n&=&-\frac{8}{3}-\frac{97}{18}\frac{1}{n}+\frac{97}{18}\frac{1}{n+1}
+\frac{29}{3}\frac{1}{n^2}-\frac{67}{3}\frac{1}{(n+1)^2}-\frac{8}{n^3}
-\frac{24}{(n+1)^3}  \nonumber \\
& &+ S_1(n) \Bigl(~\frac{67}{9}+\frac{8}{n^2}-\frac{8}{(n+1)^2} \Bigr)
        -\frac{1}{2}S'_3(\frac{n}{2}) +4\widetilde S(n)
 \nonumber \\
& &-2S'_2(\frac{n}{2}) \Bigl(S_1(n)-\frac{2}{n}+\frac{2}{n+1} \Bigr)
\nonumber \\
& & +\Bigl[1+(-1)^n \Bigr] \Biggl[~8S_2(n)\Bigl(\frac{1}{n}-\frac{1}{n+1}\Bigr)
- 2S_3(n)-4S_1(n) S_2(n) \nonumber \\
& & \qquad \qquad \qquad \quad+2S'_2(\frac{n}{2}) \Bigl(S_1(n)
-\frac{2}{n}+\frac{2}{n+1} \Bigr)  +\frac{1}{2}S'_3(\frac{n}{2})
-4\widetilde S(n)
 \nonumber \\
& & \qquad \qquad \qquad \quad+\zeta(2) \Bigl(2S_1(n)-\frac{4}{n}+\frac{4}{n+1}
 \Bigr) - \zeta(3)\Biggr]
\eea

\bigskip

\subsubsection{${\Kv}^{(1)}_{n}$}

\be
   {\Kv}^{1}_{n}=\pmatrix{K^{1,n}_{\psi},&K^{1,n}_{G},&K^{1,n}_{ NS}} ~,
\ee

with
\bea
   K^{1,n}_{\psi}&=&-3~n_f<e^2>C_F~8D^n_{\psi G}~,\\
   K^{1,n}_{G}&=&-3~n_f<e^2>C_F~8(D_{GG}^n-1)~,   \\
   K^{1,n}_{NS}&=&-3~n_f(<e^4>-<e^2>^2) C_F~8D^n_{\psi G}~.
\eea

\subsubsection{Anomalous dimensions at $n=1$ ($\overline{\rm MS}$ scheme)}

\bea
    \gamma^{0,n=1}_{NS}&=&\gamma^{0,n=1}_{\psi \psi}=0  \\
    \gamma^{0,n=1}_{\psi G}&=&0~, \\
    \gamma^{0,n=1}_{G \psi}&=&-6C_F \\
    \gamma^{0,n=1}_{G G}&=& -\frac{22}{3}C_A +\frac{8}{3}T_f =-2\beta_0   \\
    & & \nonumber  \\
    K^{0,n=1}_{NS} &=& K^{0,n=1}_{\psi}=0  \\
   & & \nonumber  \\
  \gamma^{(1),n=1}_{NS}&=& 0  \\
 & & \nonumber  \\
  \gamma^{(1),n=1}_{\psi \psi}&=& 24 C_F T_f   \\
  \gamma^{(1),n=1}_{\psi G}&=& 0  \\
  \gamma^{(1),n=1}_{G \psi}&=&
18C_F^2-\frac{142}{3}C_AC_F+\frac{8}{3}C_FT_f  \\
\gamma^{(1),n=1}_{GG}&=&8C_F T_f +\frac{40}{3}C_A T_f -\frac{68}{3}C_A^2
  =-2\beta_1 \\
  & & \nonumber  \\
  K^{(1),n=1}_{\psi} &=&K^{(1),n=1}_{G}= K^{(1),n=1}_{NS}=0
\eea

\bigskip

\newpage

\section{Coefficient Functions}

${\Cv}_{n} (C_{n}^{\gamma})$ is the coefficient function of the hadronic
(photon) operators \cite{BB}:

\be
{\Cv}_{n}(1,\bar g (Q^2)) =\pmatrix{\delta_{\psi}
     \Bigl(1+\frac{\bar g^2 (Q^2)}{16\pi^2} B^n_{\psi}  \Bigr) \cr
               \cr
      \delta_{\psi}\frac{\bar g^2 (Q^2)}{16\pi^2} B^n_{G}  \cr
           \cr
       \delta_{NS} \Bigl(1+\frac{\bar g^2 (Q^2)}{16\pi^2} B^n_{NS}  \Bigr) \cr}
\ee

\be
          C_{n}(1,\bar g (Q^2), \alpha) =\frac{e^2}{16\pi^2}
                 \delta_{\gamma} B^n_{\gamma}
\ee
and
\be
B_\gamma^n=(2/n_f)B_G^n
\ee

\subsection{$\overline{\rm MS}$ scheme \cite{MvN,V}}

\bea
B_\psi^n=B_{NS}^n&=&C_F
\left[\left(\frac{2}{n}+\frac{2}{n+1}+3\right)S_1(n-1)
+4\sum_{j=1}^{n-1}\frac{1}{j}S_1(j)-4S_2(n-1) \right.\nonumber\\
&&\left.\hspace{8cm}
 +\frac{6}{n}-9 \right] \label{bnpsi}\\
B_G^n&=&4T_f
\left[-\frac{n-1}{n(n+1)}(S_1(n)+1)-\frac{1}{n^2}+\frac{2}{n(n+1)}\right]
\eea

\subsection{Momentum subtraction \cite{KMSU,JK}}

\bea
B_\psi^n=B_{NS}^n&=&C_F
\Bigl[-2-\frac{3}{n}+\frac{8}{n+1}+\frac{4}{n^2}-\frac{4}{(n+1)^2}
\nonumber\\
&&\qquad\qquad\qquad+3S_1(n)-8S_2(n)\Bigr]
\eea

\bea
B_G^n&=&8T_f
\left[\frac{1}{n}-\frac{2}{n+1}-\frac{1}{n^2}+\frac{2}{(n+1)^2}
\right] \quad n \geq 3 \nonumber\\
&&B_G^{n=1}=0
\eea

\newpage

\section{Tensor decomposition of virtual photon-photon amplitude}

After using parity invariance, time-reversal invariance, and gauge
invariance, Brown and Muzinich~\cite{BM} have shown that
there are eight independent tensors, in other words,
eight-invariant amplitudes for virtual photon-photon
scattering. Those eight independent tensors, which are free from
kinematic singularities and kinematic zeros, are given in Eqs.(A3)-
(A10) of Ref.\cite{BM}.

Using these tensors $(I_i)_{\mu\nu\rho\tau}$~, the absorptive part of the
forward virtual photon-photon scattering amplitude $W_{\mu\nu\rho\tau}$ is
decomposed as
\be
W_{\mu\nu\rho\tau}=\sum_{i=1}^8 (I_i)_{\mu\nu\rho\tau} A_i (w, t_1, t_2)
\ee
where the $A_i$ are the invariant amplitudes and
\be
  w=p\cdot q,  \quad t_1=q^2=-Q^2, \quad t_2=p^2=-P^2~.
\ee
In order to implement crossing symmetry under
$q \rightarrow -q$ and $\mu \leftrightarrow \nu$, we form the
even combinations, $ I_1$, $I_2+I_3$, $I_4$, $I_5$, $I_7+I_8$,
and the odd combinations, $I_2-I_3$, $I_7-I_8$, $I'_6=2I_6-3wI_7-wI_8+(t_1
t_2/w)(I_2-I_3)$.  It is noted that the combinations $I_2-I_3$ and $I_7-I_8$
are anti-symmetric under the interchange of $\mu \leftrightarrow \nu$ and
$\rho \leftrightarrow \tau$, while the rest of the combinations are
symmetric.  In terms of these crossing-even  and -odd combinations,
the amplitude $ W=\sum^8_{i=1}I_i A_i $ is rearranged as follows:
\bea
\sum^8_{i=1}I_i A_i&=&I_1 A_1+\frac{1}{2}(I_2+I_3)(A_2+A_3 )+I_4 A_4+I_5 A_5
\nonumber\\
&+& \frac{1}{2}(I_7+I_8) (A_7+A_8 +2 w A_6) +\frac{1}{2}I'_6 A_6
\nonumber\\
&+& \frac{1}{2}(I_2-I_3) (A_2-A_3-\frac{t_1 t_2}{w}A_6 )+
       \frac{1}{2}(I_7-I_8) (A_7-A_8 + w A_6)
\eea
Now $g_1^\gamma=2W_4^{\gamma}$ is written in terms of invariant amplitudes
as
\bea
   g_1^\gamma &\propto& a_{1111}-a_{1-11-1} \\
       &=&w^2 (A_2-A_3-\frac{t_1 t_2}{w}A_6) -t_1 t_2 (A_7-A_8 + w A_6)
\eea
which is obtained from Eq.(A14) of Ref.\cite{BM}. Here $a_{1111}$
($a_{1-11-1}$) represents the $s$-channel helicity amplitude for
$(+1)\gamma + (\pm1)\gamma \longrightarrow (+1)\gamma + (\pm1)\gamma$~.
It is noted that
$(A_2-A_3-(t_1 t_2/w)A_6)$ is the invariant amplitude associated with
$(I_2-I_3)$ , while
$(A_7-A_8 + w A_6)$  is associated with $(I_7-I_8)$.
In the limit $t_2=p^2=0$ or in the case that
$w=p\cdot q, t_1=q^2 \gg t_2=p^2 $, the second term $t_1 t_2
(A_7-A_8 + w A_6)$  does not contribute to $g_1^{\gamma}$.

In fact we observe that the tensor $(I_2-I_3)\equiv I_{-}$ is
associated to $g_1^{\gamma}$ while  $(I_7-I_8)\equiv J_{-}$ is
associated  to $g_2^{\gamma}$. It can be shown that
\be
\epsilon_{\mu\nu\lambda\sigma}q^{\lambda}
{\epsilon_{\rho\tau}}^{\sigma\beta}p_{\beta}
= \frac{1}{p\cdot q} I_-~,
\ee
and in the limit of $-q^2, p\cdot q \gg -p^2$
\be
\Bigl[\epsilon_{\mu\alpha\lambda\sigma}q_{\nu}q^{\alpha}-
    \epsilon_{\nu\alpha\lambda\sigma}q_{\mu}q^{\alpha} -
   \epsilon_{\mu\nu\lambda\sigma} \Bigr]
{\epsilon_{\rho\tau}}^{\sigma\beta}p_{\beta}p^{\lambda}
=J_{-}~.
\ee
Now using an identity
\begin{equation}
g_{\mu\nu}\epsilon_{\alpha\beta\gamma\delta}=
g_{\mu\alpha}\epsilon_{\nu\beta\gamma\delta}+
g_{\mu\beta}\epsilon_{\alpha\nu\gamma\delta}+
g_{\mu\gamma}\epsilon_{\alpha\beta\nu\delta}+
g_{\mu\delta}\epsilon_{\alpha\beta\gamma\nu}
\end{equation}
we get
\begin{eqnarray}
&&\epsilon_{\mu\nu\lambda\sigma}q^\lambda
(p\cdot q\ {\epsilon_{\rho\tau}}^{\sigma\beta}p_\beta-\epsilon_{\rho\tau
\alpha\beta}p^\beta p^\sigma q^\alpha)\nonumber\\
&=&-\left[
\epsilon_{\mu\alpha\lambda\sigma}q_\nu q^\alpha -
\epsilon_{\nu\alpha\lambda\sigma}q_\mu q^\alpha -
\epsilon_{\mu\nu\lambda\sigma}q^2
\right]{\epsilon_{\rho\tau}}^{\alpha\beta}p_\beta p^\lambda
\end{eqnarray}
Hence we have from Eq.(2.3)
\begin{equation}
W_{\mu\nu\rho\tau}=\frac{1}{(p\cdot q)^2}[(I_{-})_{\mu\nu\rho\tau}\
g_1^{\gamma} -(J_{-})_{\mu\nu\rho\tau}\ g_2^{\gamma}].
\end{equation}

Finally it is interesting to see the relation between the polarized
photon structure functions $g_1^{\gamma}$ and $g_2^{\gamma}$ and polarized
nucleon structure functions $g_1$ and $g_2$. By introducing the polarization
vectors, $\epsilon^{*\rho}$ and
$\epsilon^\tau$ for the target photon just like those for the
gluon target discussed by Gabrieli and Ridolfi \cite{GRI}, we have
\begin{eqnarray}
iW_{\mu\nu}^A&=&\epsilon^{*\rho}W_{\mu\nu\rho\tau}\epsilon^\tau
=W_{\mu\nu\rho\tau}\frac{1}{2}(\epsilon^{*\rho}\epsilon^{\tau}-
\epsilon^{*\tau}\epsilon^{\rho}) \nonumber\\
&=&
W_{\mu\nu\rho\tau}(-\frac{i}{2\sqrt{|p^2|}}\epsilon^{\rho\tau
\gamma\delta}p_\gamma s_\delta)~.
\end{eqnarray}
where $s$ is the longitudinal spin vector for the target photon.
After using the relation $p\cdot s=0$, we get
\begin{equation}
W_{\mu\nu}^A=\frac{\sqrt{|p^2|}}{p\cdot q}\left[
\epsilon_{\mu\nu\lambda\sigma}q^\lambda s^\sigma g_1^{\gamma}
+\epsilon_{\mu\nu\lambda\sigma}q^\lambda(p\cdot q s^\sigma-
q\cdot sp^\sigma)\frac{g_2^{\gamma}}{p\cdot q}\right]
\end{equation}
which, apart from the factor $\sqrt{|p^2|}$, has exactly the same form as
Eq.(2.4) of Kodaira et al.~\cite{KMSU} which defines the polarized nucleon
structure functions $g_1$ and $g_2$, and also as Eq.(9) in Ref.~\cite{GRI}.

\newpage

\newpage
\vspace{3cm}
\noindent
{\large Figure Captions}
\baselineskip 16pt

\begin{enumerate}

\item[Fig. 1] \quad
Deep inelastic scattering on a polarized virtual photon
in polarized $e^{+}e^{-}$ collision,
$e^{+}e^{-} \rightarrow  e^{+}e^{-} + $ hadrons (quarks and gluons).
The arrows indicate the polarizations of the $e^{+}$, $e^{-}$ and
virtual photons.
The mass squared of the \lq\lq probe \rq\rq (\lq\lq target \rq\rq)
photon is $-Q^2(-P^2)$ ($\Lambda^2 \ll P^2 \ll Q^2$).

\item[Fig. 2] \quad
Forward scattering of a virtual photon with momentum $q$ and
another virtual photon with momentum $p$. The Lorentz indices are
denoted by $\mu,\nu,\rho,\tau$.

\item[Fig. 3] \quad
Polarized virtual photon structure function $g_1^\gamma(x,Q^2,P^2)$
to the next-to-leading order (NLO) in units of $(3\alpha n_f <e^4>/\pi)
\ln(Q^2/P^2)$ for $Q^2=30$ GeV$^2$, and $P^2=1$ GeV$^2$ 
and the QCD scale parameter $\Lambda=0.2$ GeV with $n_f=3$ (solid line).
We also plot the leading order (LO) result (long-dashed line),
the Box (tree) diagram (2dash-dotted line) and the Box including
non-leading contribution, Box (NL) (short-dashed line).

\item[Fig. 4] \quad
Virtual photon structure function $g_1^\gamma(x,Q^2,P^2)$
for $Q^2=100$ GeV$^2$, and $P^2=1$ GeV$^2$ with  $\Lambda=0.2$ GeV, $n_f=3$.

\item[Fig. 5] \quad
Virtual photon structure function $g_1^\gamma(x,Q^2,P^2)$
for $Q^2=30$ GeV$^2$, and $P^2=3$ GeV$^2$ with  $\Lambda=0.2$ GeV, $n_f=3$.

\item[Fig. 6] \quad
Point-like piece of the real photon structure function
$g_1^\gamma(x,Q^2)$ in NLO
for $Q^2=30$ GeV$^2$ with  $\Lambda=0.2$ GeV, $n_f=3$ (solid line).
Also plotted are the LO result (long-dashed line) and the Box (tree)
diagram contribution (short-dashed line).

\end{enumerate}

\newpage
\pagestyle{empty}
\input epsf.sty
\begin{figure}
\centerline{
\epsfxsize=12cm
\epsfbox{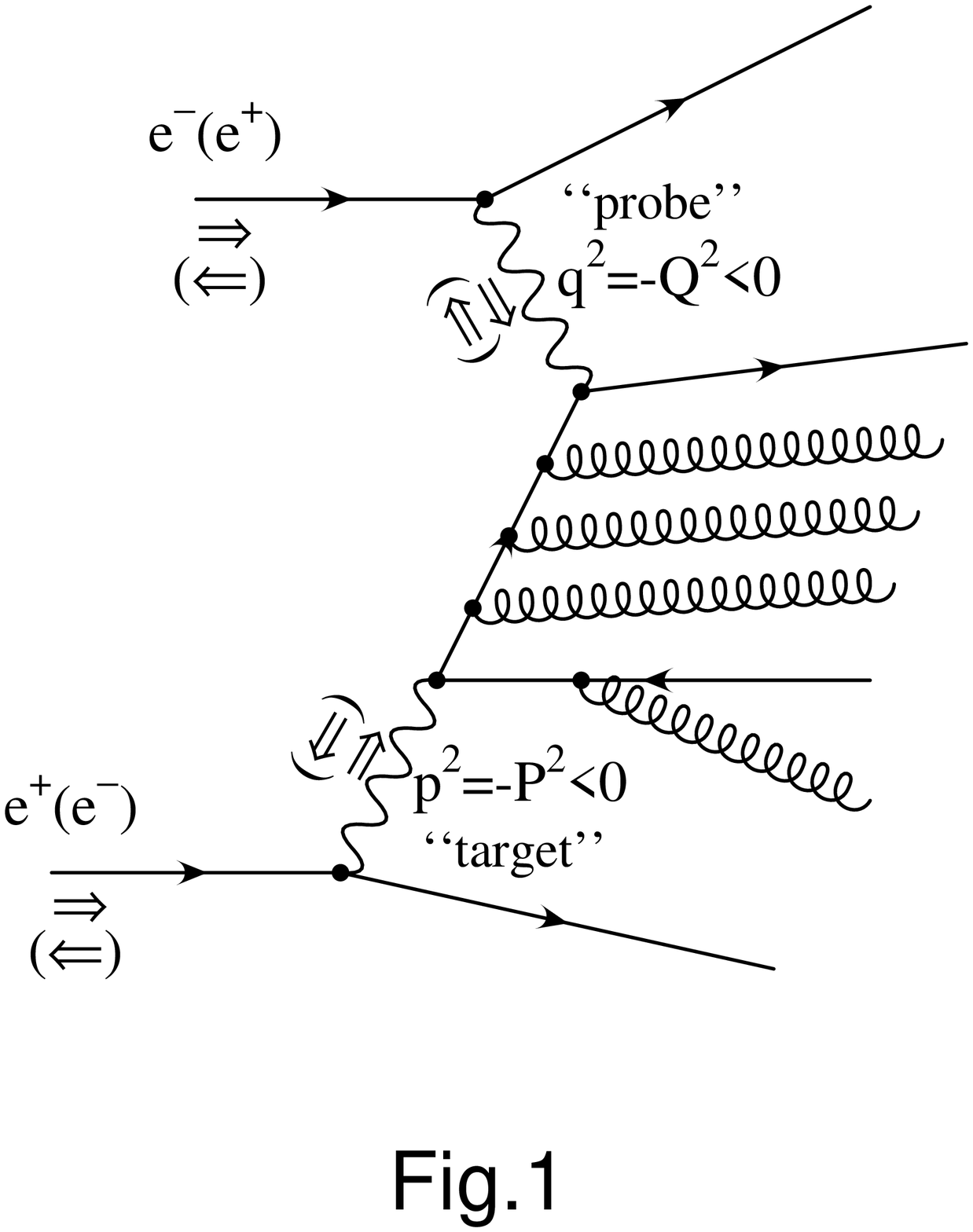}}
\end{figure}

\newpage
\pagestyle{empty}
\input epsf.sty
\begin{figure}
\centerline{
\epsfxsize=8cm
\epsfbox{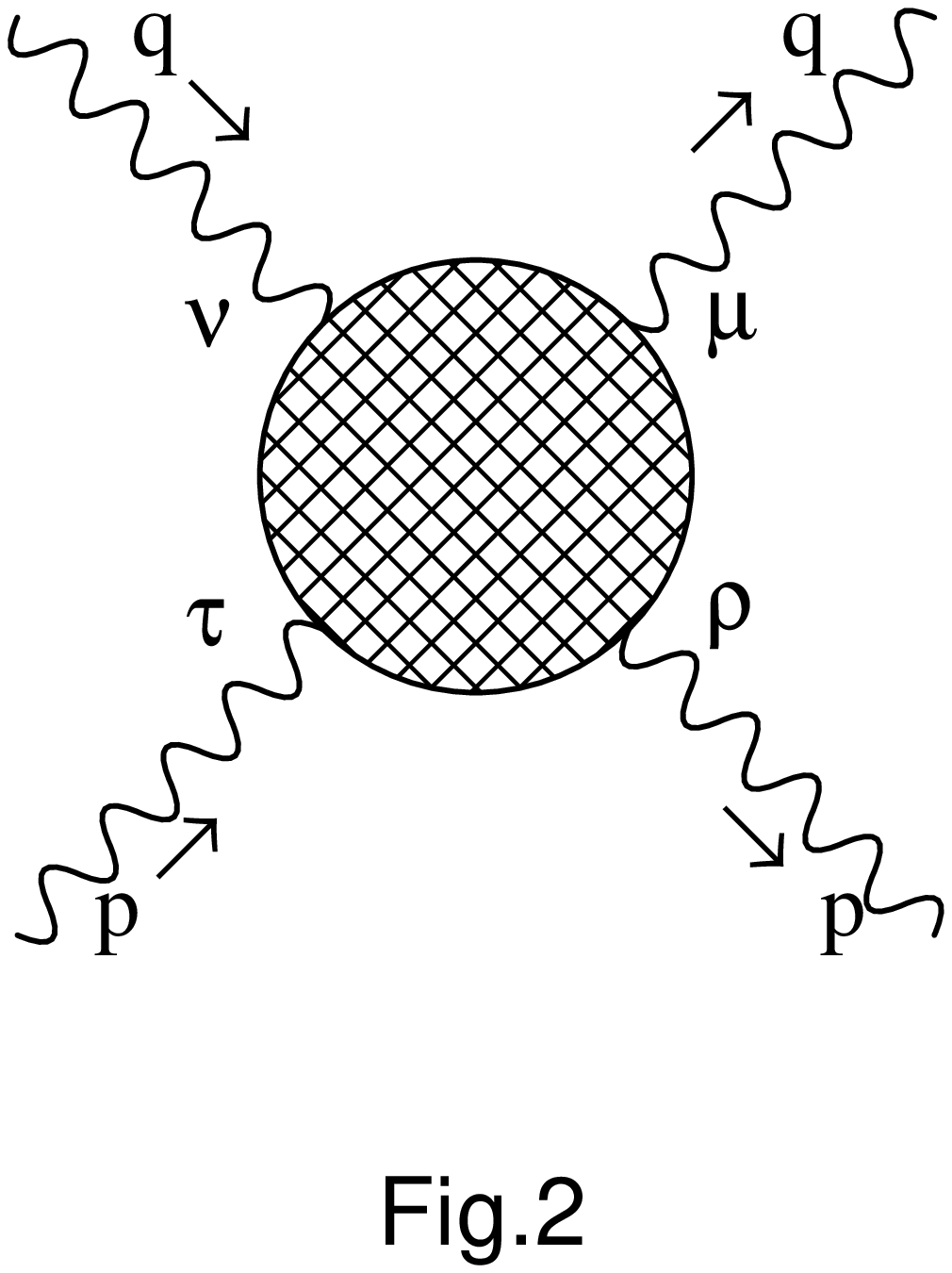}}
\end{figure}

\newpage
\pagestyle{empty}
\input epsf.sty
\begin{figure}
\centerline{
\epsfxsize=14cm
\epsfbox{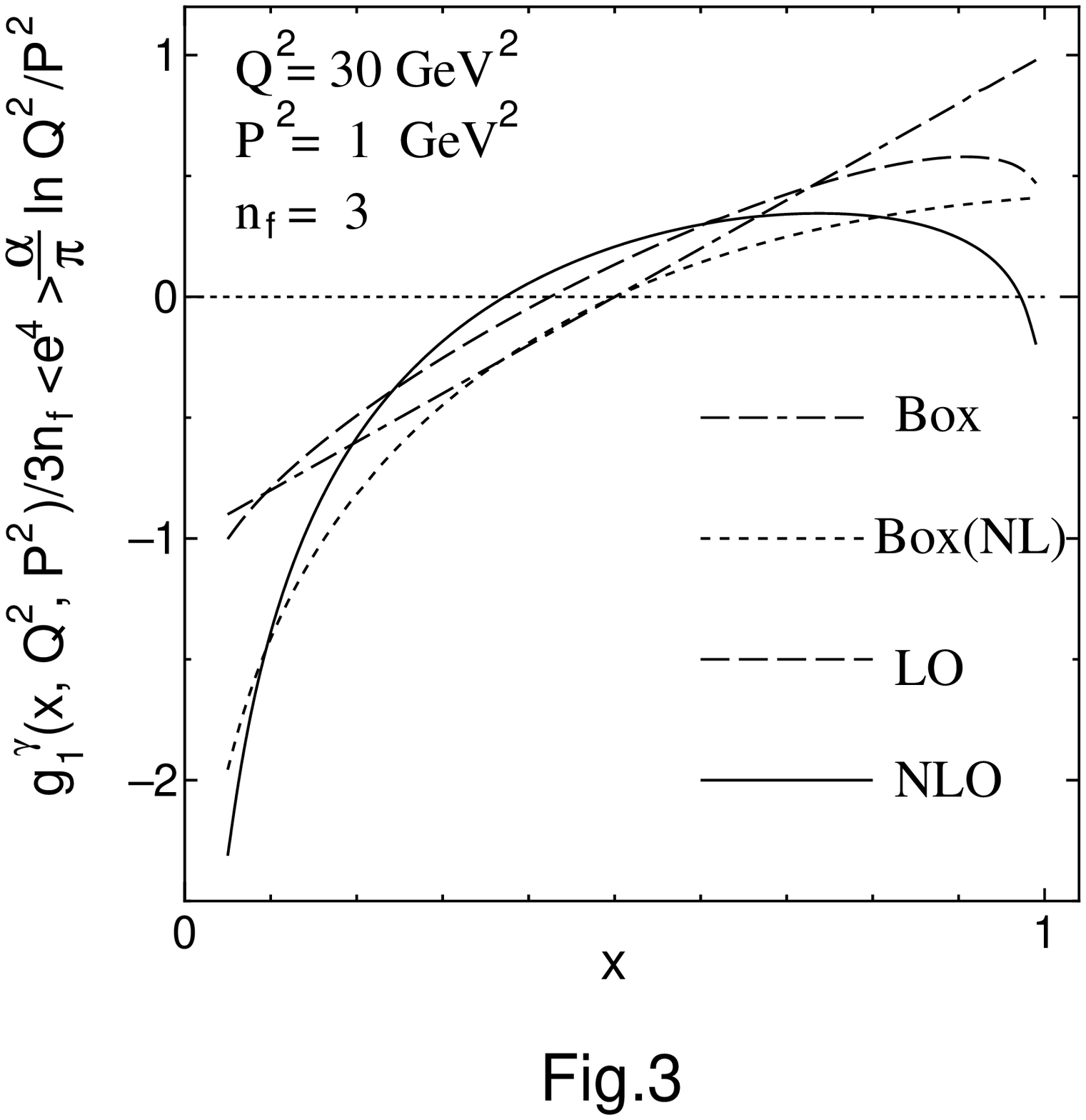}}
\end{figure}

\newpage
\pagestyle{empty}
\input epsf.sty
\begin{figure}
\centerline{
\epsfxsize=14cm
\epsfbox{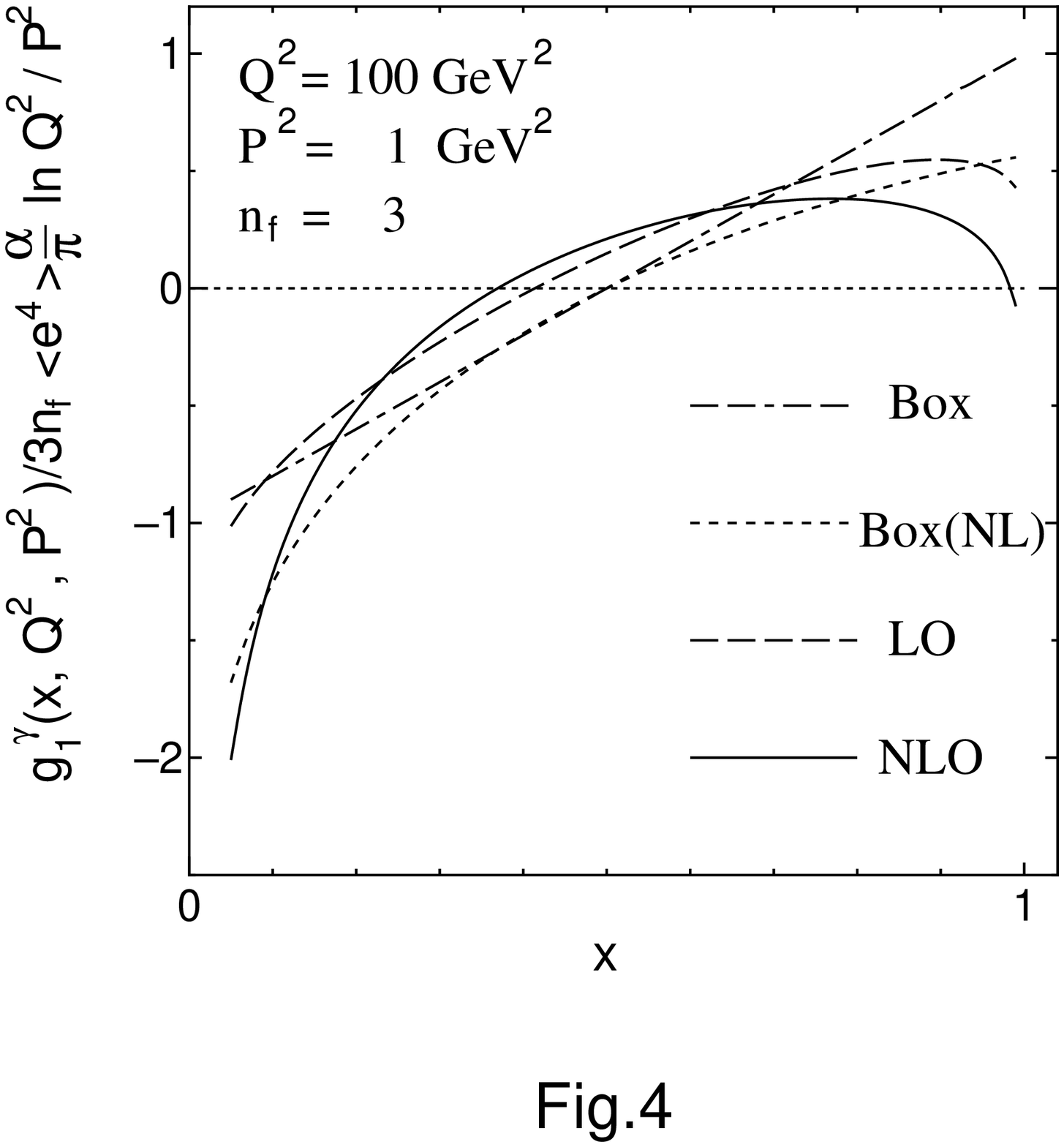}}
\end{figure}

\newpage
\pagestyle{empty}
\input epsf.sty
\begin{figure}
\centerline{
\epsfxsize=14cm
\epsfbox{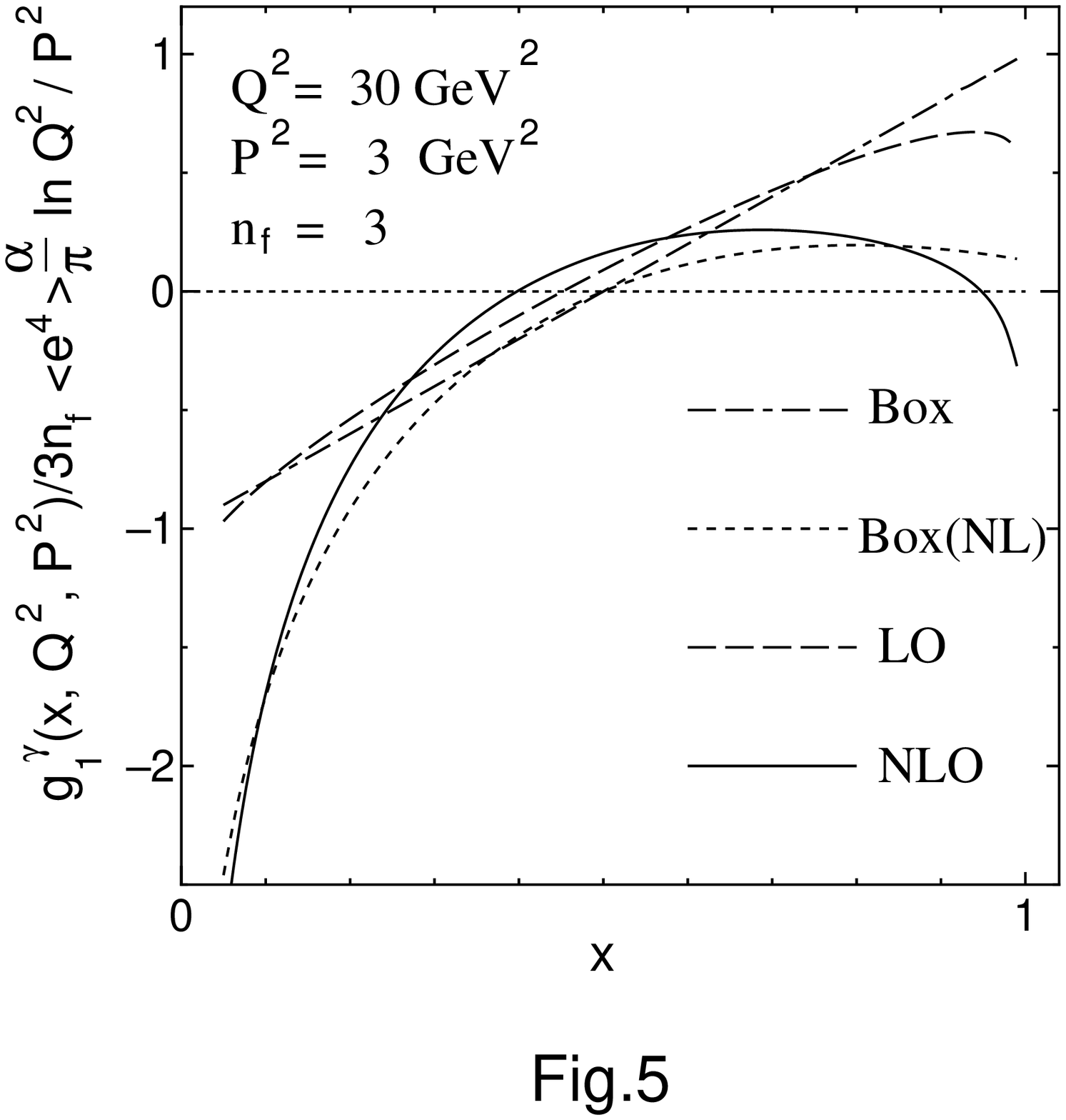}}
\end{figure}

\newpage
\pagestyle{empty}
\input epsf.sty
\begin{figure}
\centerline{
\epsfxsize=14cm
\epsfbox{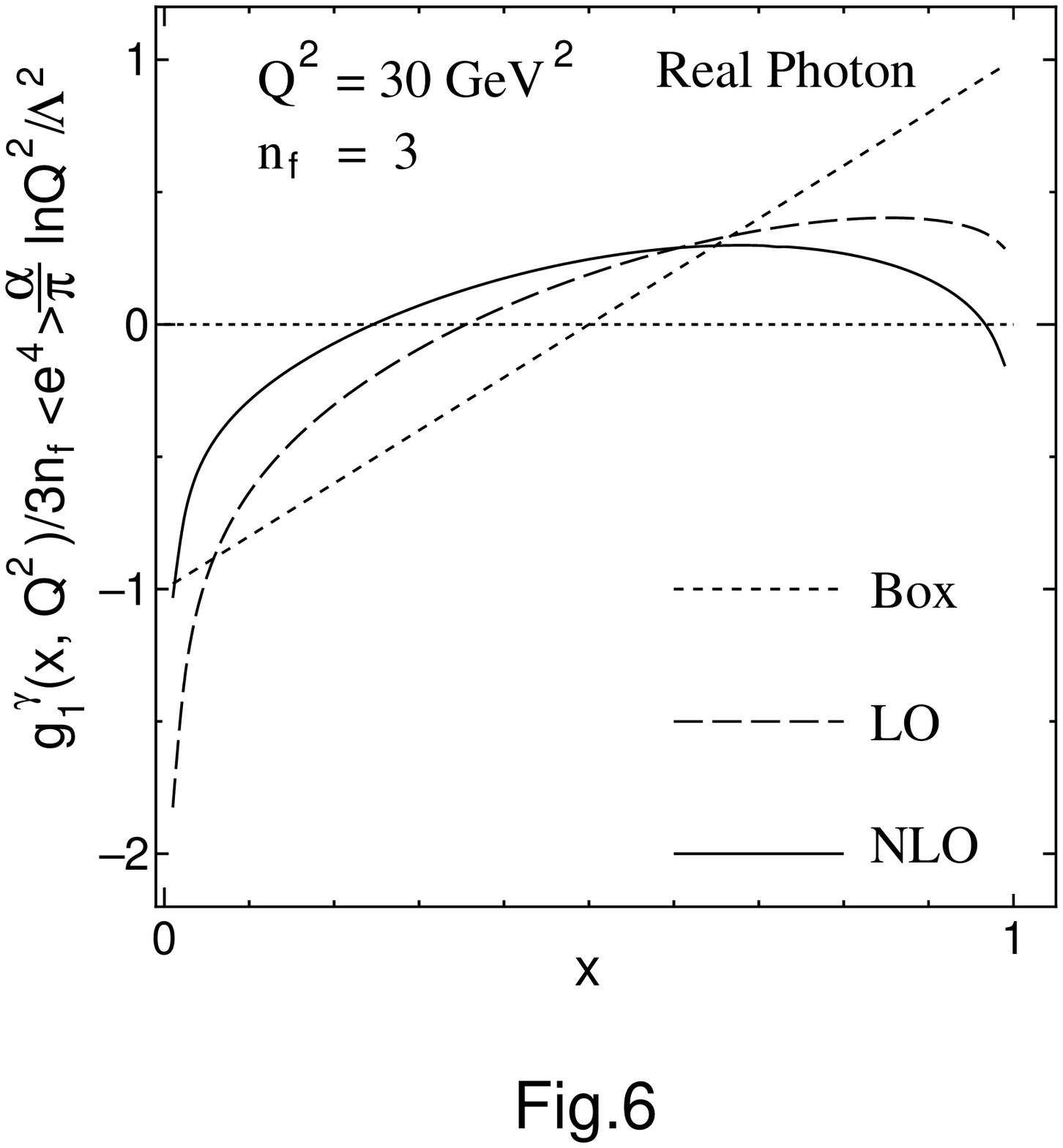}}
\end{figure}

\end{document}